\documentclass[10pt,longbibliography,physrev,aps,twocolumn]{revtex4-2}
\usepackage{graphicx,epsfig}
\usepackage{times}
\usepackage{amsmath,amssymb,wasysym}
\usepackage{dsfont} % \mathds
\usepackage{xcolor}
\usepackage[linktocpage=true,colorlinks=true, pdfborder={0 0 0},linkcolor=blue,citecolor=blue,urlcolor=blue,anchorcolor=black,linkcolor=blue]{hyperref}

\newcommand{\rem}[1]{}
\newcommand{\bra}[1]{\langle#1|}
\newcommand{\ket}[1]{|#1\rangle}
\newcommand{\braket}[2]{\langle#1|#2\rangle}
\newcommand{\ident}{\mathds{1}}
\newcommand{\op}[1]{\hat{#1}}
\newcommand{\Ham}{H}
\newcommand{\ev}{E}

\newcommand{\evEP}{\ev_{\text{EP}}}
\newcommand{\order}{n}
\newcommand{\Hp}{H_1}
\newcommand{\hp}{h_1}
\newcommand{\GF}{G}

\newcommand{\rca}{\xi}
\newcommand{\rcaintern}{\xi^{(R,L)}}
\newcommand{\PF}{K}

\newcommand{\projector}{\op{P}}
\newcommand{\nilpotent}{N}
\newcommand{\REP}{R_{\text{EP}}}
\newcommand{\LEP}{L_{\text{EP}}}

\newcommand{\Ha}{H_{\text{a}}}
\newcommand{\Hb}{H_{\text{b}}}
\newcommand{\Ea}{E_{\text{a}}}
\newcommand{\Eb}{E_{\text{b}}}
\newcommand{\cab}{V}
\newcommand{\cba}{W}

\newcommand{\rEPb}{r_{\text{EP,b}}}
\newcommand{\Rv}{R}
\newcommand{\rva}{r_{\text{a}}}
\newcommand{\rvb}{r_{\text{b}}}

\newcommand{\lEPb}{l_{\text{EP,b}}}
\newcommand{\Lv}{L}

\newcommand{\lvb}{l_{\text{b}}}
\newcommand{\Vector}{\Psi}
\newcommand{\vectora}{\psi_{\text{a}}}
\newcommand{\vectorb}{\psi_{\text{b}}}
\newcommand{\ham}{h}
\newcommand{\GFa}{G_{\text{a}}}
\newcommand{\GFb}{G_{\text{b}}}
\newcommand{\Gaa}{G_{\text{aa}}}
\newcommand{\Gab}{G_{\text{ab}}}
\newcommand{\Gba}{G_{\text{ba}}}
\newcommand{\Gbb}{G_{\text{bb}}}
\newcommand{\projectorb}{p_{\text{b}}}
\newcommand{\ind}{l}
\newcommand{\quoting}[1]{``#1''}
\newcommand{\transpose}{{\text{T}}}

\newcommand{\normV}[1]{||{#1}||_2}

\newcommand{\normspec}[1]{||{#1}||_{\text{2}}}
\newcommand{\normFro}[1]{||#1||_{\text{F}}}
\newcommand{\pf}{\omega}

\newcommand{\freqEP}{\omega_{\text{EP}}}

\begin{document}

\title{Generalized Petermann factor of non-Hermitian systems at exceptional points}
\author{Julius Kullig}
\author{Jan Wiersig}
\affiliation{Institut f{\"u}r Physik, Otto-von-Guericke-Universit{\"a}t Magdeburg, Postfach 4120, D-39016 Magdeburg, Germany}
\email{jan.wiersig@ovgu.de}
\author{Henning Schomerus}
\email{h.schomerus@lancaster.ac.uk}
\affiliation{Department of Physics, Lancaster University, Lancaster, LA1 4YB, United Kingdom}
\date{\today}
\begin{abstract}
The nonorthogonality of modes in open systems significantly modifies their resonant response, resulting in quantitative  and qualitative deviations from Breit-Wigner resonance relations. For isolated resonances with a Lorentzian lineshape, the deviations amount to a quantitative enhancement of the resonance linewidth, given by the Petermann factor, which is determined by the overlap of left and right eigenmodes of the underlying effectively non-Hermitian Hamiltonian.
The Petermann factor diverges at exceptional points (EPs), where complex resonance frequencies become degenerate, and right and left eigenmodes are orthogonal to each other. This divergence signifies a qualitative departure from a Lorentzian lineshape, which has moved into the focus of recent attention. 
In this work, we develop the analog of the Petermann factor for this qualitatively modified response at an EP, and describe how this EP Petermann factor manifests in a variety of physical scenarios.
First, we identify this analog in physical terms as an enhancement of the response of a system to external or parametric perturbations. Utilizing two  natural orthogonally projected reference systems based on the right and left eigenvectors, we show that each choice carries a precise geometric interpretation that naturally extends the notion of the Petermann factor for isolated resonances to EPs. The two choices can be combined into an overall EP Petermann factor, which again can be expressed in purely geometric terms. 
Second, we illuminate the geometric mechanisms that determine the size of the EP Petermann factor, by considering the role of modes participating in the degeneracy and those that remain spectrally separated from it. This also leads to a systematic description to evaluate this factor.
Third, we design a system that allows us to study the EP Petermann factor in a specific physical setup, consisting of two microrings coupled to a waveguide with embedded semitransparent mirrors. With this example, we demonstrate that our approach gives more accurate results for the spectral response strength than conventional truncation schemes.
These results complete the description of systems operating at exceptional points in the same way as the original Petermann factor does for isolated resonances, and pave the way
to the design of open systems with unconventional spectral response, be it in emission or in response to static and dynamic perturbations.
\end{abstract}
\maketitle

\section{Introduction}
% general
The field of non-Hermitian physics has attracted a considerable amount of attention in various subdisciplines, such as photonics~\cite{EMK18,ORN19}, acoustics~\cite{HHS24}, and condensed-matter physics~\cite{BBK21}. The term \quoting{non-Hermitian} refers to the (effective) Hamiltonian~\cite{Feshbach58,Feshbach62}, which describes not only conservative but also dissipative dynamics of waves, stemming, e.g., from material gain, material loss, and radiation loss. 

% nonorthogonality
The non-Hermiticity can lead to complex-valued energy eigenvalues, where the imaginary part describes decay or growth. A more intriguing effect is the eigenmode nonorthogonality, which has several interesting physical implications such as nonexponential transient decay~\cite{TE05} and amplification~\cite{MGT14}, power oscillations in optical waveguides~\cite{MEC08,RMG10}, chirality in perturbed whispering-gallery microcavities~\cite{WES11,Wiersig11}, sensitivity of resonance widths under perturbation~\cite{FS12}, limitation of mode selectivity~\cite{DG18}, 
directional amplification \cite{HS20}, adiabatic amplification \cite{OS25},
and quantum excess noise in lasers~\cite{Petermann79}. 
% Petermann factor 
The latter shows up experimentally as a laser linewidth broadening which is quantified by the Petermann factor~\cite{Petermann79,Siegman86,SFP00,Schomerus09,CS12,PNS14}. The Petermann factor~$K$ is not only a linewidth enhancement factor in the laser context but is a general measure of the mode nonorthogonality. As such it quantifies the enhanced response of a given eigenstate to perturbations and noise~\cite{HS20}, which is relevant for non-Hermitian topological sensors~\cite{BB20}. 

% EPs
The Petermann factor diverges to infinity~\cite{WBW96,Berry03,LRS08} at so-called exceptional points (EPs) in parameter space~\cite{Kato66}. At such an EP of order $\order$, exactly $\order$ eigenstates and their corresponding eigenvalues of the Hamiltonian coalesce. EPs are highly sensitive to perturbations: A perturbation of  strength~$\varepsilon > 0$ results generically in eigenvalue changes proportional to the $\order$th root of~$\varepsilon$~\cite{Kato66}. This sensitivity makes EPs good candidates for sensing applications~\cite{Wiersig14b,COZ17,HHW17,XLK19,Wiersig20b,Wiersig20c,KCE22,BKB24}. The response of the system at the EP in terms of eigenvalue changes can be quantified by the spectral response strength~$\rca$~\cite{Wiersig22}.

% aim 
The aim of this paper is to extend the concept of the Petermann factor, which was originally introduced for isolated resonances, to the degenerate situation at an EP. Physically, this \emph{EP Petermann factor} quantifies the enhancement of the response strength above that of reference systems from which the mode-nonorthogonality is removed. For these systems, there are two natural choices: One pertains to the excitation of the system, while the other relates to the detection of the resulting response.
The corresponding enhancement factors carry a precise geometric interpretation that naturally extends the notion of the Petermann factor for isolated resonances to EPs, and link the physical enhancement of the response strength to purely geometric information quantifying nonorthogonality of  the right and left generalized eigenvectors.
This information can further be combined into an overall EP Petermann factor, which again carries a precise geometric interpretation quantifying mode non-orthogonality.
To substantiate the physical interpretation of these concepts, we illuminate the role of the EP Petermann factor in various contexts, such as system response to perturbations and quantum noise.
In addition to these theoretical insights, we present a detailed prescription for determining the EP Petermann factor from both analytical and numerical models.
Finally, we demonstrate the practical importance of the EP Petermann factor by comparing its predictions for the spectral response strength to those of standard truncation schemes. Through both an analytical toy model and a full numerical simulation of a realistic photonic system, we reveal how the inclusion of the EP Petermann factor leads to significantly more accurate results.

% outline 
The structure of the paper is as follows. Section~\ref{sec:background} presents the background and context to these questions. Section~\ref{sec:response} introduces the EP Petermann factor physically in terms of the spectral response strength and identifies its geometric interpretation in terms of mode-nonorthogonality, while Sec.~\ref{eq:noise} describes how the very same notions carry over to quantum-limited excess noise. Detailed prescriptions for evaluation of the EP Petermann factor are described in Sec.~\ref{sec:detailed}. Results for a realistic model system consisting of two microrings coupled via a waveguide are presented in Sec.~\ref{sec:microrings}. The conclusions are given in Sec.~\ref{sec:conclusion}.

\section{Background and objective}
\label{sec:background}
\subsection{Definitions and preliminary considerations}
In this work, we consider the consequences of mode nonorthogonality for the physical response of systems described by effectively non-Hermitian Hamiltonians $\op{H}$ in different spectral scenarios. We do not make any specific assumptions about the nature of the effective Hamiltonian, i.e., whether it arises from a mean-field description, post selection, or elimination of unobserved parts or freedoms, but assume for concreteness that the relevant part of the point spectrum can be described by an $N\times N$-dimensional matrix $H$. In a given orthonormal basis $\{|n\rangle\}_{1\leq n\leq N}$,  $\langle n|m\rangle=\delta_{nm}$, the 
elements of this matrix are given by $H_{nm}=\langle n|\op{H}|m\rangle$. We denote the identity operator as $\op{I}$, and the $N\times N$-dimensional identity matrix as $\openone$.

The spectrum is then determined by the eigenvalue equations
\begin{equation}
  \op{H}\ket{R_l}=E_l\ket{R_l},\quad   \bra{L_l} \op{H}=E_l \bra{L_l},
\label{eq:eigenv}  
\end{equation}
where $E_l$ are the eigenvalues and
 $\ket{R_l}$ and $\bra{L_l}$ are the right and left eigenvectors.
The eigenvalues can also be obtained from the secular equation
\begin{equation}
\mathrm{det}(E\openone-H)=0.
\label{eq:seceq}
\end{equation}
The nature of this eigenvalue problem depends on whether the Hamiltonian is Hermitian or not, and whether the eigenvalue spectrum displays degeneracies. In operator form, the Hamiltonian is Hermitian (equivalently, self-adjoint) if it fulfills $\langle \psi|\op{H} \varphi\rangle=\langle \op{H}\psi| \varphi\rangle$
for any states $\ket{\psi}$ and $\ket{\varphi}$, hence, $\op{H}=\op{H}^\dagger$,  where the dagger denotes Hermitian conjugation. In matrix form, this entails $H_{nm}=H_{mn}^*$. In this background section we briefly review the key aspects of these distinctions that serve as the starting point for our considerations.

\subsection{Spectral decomposition of Hermitian matrices}

We recall that for a Hermitian Hamiltonian, the eigenvalues $E_l$ are real, while the right and left eigenvector can be identified via $\bra{L_l}=\ket{R_l}^\dagger$. Additionally, the eigenvectors of distinct eigenvalues $E_k\neq E_l$ are  mutually orthogonal, $\langle R_k|R_l\rangle=0$, and individual eigenvectors can be normalized to obey  $\langle R_l|R_l\rangle=1$.
Furthermore,  this construction can be straightforwardly  extended to the case of degenerate eigenvalues. The mathematical reason is that in Hermitian systems, the algebraic multiplicity $n_l$, given by the multiplicity of roots in the secular equation \eqref{eq:seceq}, is identical to the geometric multiplicity $g_l$, given by the number of linearly independent eigenvectors admitted by Eq.~\eqref{eq:eigenv}. Employing a suitable orthogonal basis in the subspace spanned by these vectors, we therefore can construct an orthonormal basis for the complete system, with
\begin{equation}
\langle R_k|R_l\rangle=\delta_{kl} \quad\mbox{(orthonormal basis)}.
\end{equation}
These features entail that a Hermitian matrix can be diagonalized by a unitary transformation,
$H=U\,\mathrm{diag}(E_l)\,U^\dagger$, where $U_{ml}=\langle m\ket{R_l}$ in terms of the basis $\ket{m}$ in which $\op{H}$ was originally formulated.
This corresponds to the spectral decomposition
\begin{equation}
    \op{H}=\sum_l E_l\ket{R_l}\bra{R_l}\quad\mbox{(Hermitian matrix)},
\end{equation}
where the sum runs over all states of the orthonormal eigenbasis, so that degenerate eigenvalues are repeated $n_l$ times.

For our purposes, it will be useful to introduce the projectors $\hat P_l$ onto the eigenspaces of distinct eigenvalues. For nondegenerate eigenvalues, $\hat P_l=\ket{R_l}\bra{R_l}$, while generally
\begin{equation}
    \hat P_l=\sum_{m|E_m=E_l} \ket{R_m}\bra{R_m}\quad\mbox{(Hermitian matrix)},
\end{equation}
where the participating eigenvectors are again suitably ortho\-normalized.
In contrast to the remaining freedom in the exact basis choice (including the arbitrary phase choice for eigenvectors of isolated eigenvalues), these projectors are uniquely defined.
Furthermore, each projector is  Hermitian (such projectors are known as orthogonal projectors), and their collection fulfills
\begin{equation}
\hat P_l\hat P_k=\hat P_k\hat P_l=\delta_{kl}\hat P_l.
\label{eq:projcond}
\end{equation}
The spectra decomposition then takes the compact form
\begin{equation}
    \op{H}=\sum_l E_l  \hat P_l\quad\mbox{(Hermitian matrix)},
\end{equation}
where the sum over $l$ is now restricted to distinct eigenvalues.

\subsection{Spectral decomposition of non-Hermitian matrices with non-degenerate eigenvalues}

A very similar construction can be carried out for non-Hermitian matrices in which all eigenvalues are non-degenerate.
The main differences are that the eigenvalues~$E_l$ are generally complex, that the right eigenvectors are no longer mutually orthogonal (the same applies to the left eigenvectors), and that the right and left eigenvectors are no longer linked by Hermitian conjugation.
Instead, with suitable normalization, these sets of eigenvectors form a biorthogonal basis, with 
\begin{equation}
\langle L_k|R_l\rangle=\delta_{kl}.
\label{eq:biorth}
\end{equation}
We remark that this still leaves some freedom to employ different normalization conventions.   

For such non-degenerate systems, this biorthogonal basis can be used to diagonalize the Hamiltonian by a similarity transformation
\begin{equation}
    H=V\,\mathrm{diag}(E_l)\,V^{-1},
    \label{eq:diagnonh}
\end{equation}
where $V_{ml}=\langle m\ket{R_l}$, while $(V^{-1})_{lm}=\langle L_l\ket{m}$. Therefore, the matrix $V$ is generally no longer unitary, but remains invertible. 
The corresponding spectral decomposition takes the form
\begin{equation}
   \hat H=\sum_l E_l\hat P_l, \quad \hat P_l=
   %\frac{1}{\langle L_l|R_l\rangle}
   \ket{R_l}\bra{L_l},
   \label{eq:decnonh1}
\end{equation}
where in the presently considered non-degenerate case the projectors $\hat P_l $ are all of rank 1. These projectors still fulfill  Eq.~\eqref{eq:projcond}, but they are generally no longer Hermitian. Such projectors are called \textit{oblique}.

\subsection{Exceptional points}
Degeneracies of non-Hermitian systems differ from degeneracies in Hermitian systems in that along with the eigenvalues, the eigenvectors of the participating eigenvalues can also coalesce. For typical degeneracies in non-Hermitian systems, the geometric multiplicity is indeed smaller than the algebraic multiplicity, and in the generic case (i.e., without further fine tuning of parameters), the geometric multiplicity is one. These generic non-Hermitian degeneracies are called EPs.
To cover these cases, we can view the indices~$l$ as labels of separate eigenvectors, each associated with $n_l$ degenerate eigenvalues. This includes the case that the spectrum displays several EPs, while scenarios with higher geometric multiplicity can, in the context of the present work, be considered as additional accidental degeneracies between different EPs.

As the number of independent eigenvectors is reduced, these vectors do not form a complete basis (the Hamiltonian is then said to be defective). An important consequence is that the Hamiltonian can no longer be diagonalized. Instead, a complete basis can be obtained by supplementing each right eigenvector $|R_l\rangle$ by a Jordan chain of $n_l-1$ covectors $|J^{(l)}_n\rangle$,  fulfilling
\begin{equation}
    (\op{H}-E_l\op{I})|J^{(l)}_n\rangle=|J^{(l)}_{n-1}\rangle,\quad n=2,3,\ldots,n_l
\end{equation}
where nominally 
$|J^{(l)}_1\rangle=|R_l\rangle$.
Analogously, one can define a chain of left covectors
$\langle \tilde J^{(l)}_m|$, which fulfill 
\begin{equation}
   \langle \tilde J^{(l)}_m| (\op{H}-E_l\op{I})=\langle \tilde J^{(l)}_{m+1}|,\quad m=1,\ldots,n_l-1,
\end{equation}
where now nominally 
$\langle \tilde J^{(l)}_{n_l}|=\langle L_l|$.
With this labeling convention, we can furthermore introduce the generalized biorthogonality condition 
\begin{equation}
    \langle \tilde J^{(l)}_m|J^{(l)}_n\rangle=\delta_{mn}.
    \label{eq:jordanbiorth}
\end{equation}
This biorthogonality condition will be used at EPs, where it replaces the conventional biorthogonality condition in Eq.~(\ref{eq:biorth}). This universal rule applies throughout the remainder of the paper.
The transformation into this biorthogonal basis corresponds to a Jordan decomposition
\begin{equation}
    H=TJT^{-1},
    \label{eq:jordandecomp}
\end{equation}
where $T$ is a generally non-unitary similarity transformation, while $J$ is the Jordan normal form, consisting of $n_l$-dimensional blocks 
\begin{equation}
J_l=E_l\openone+D,
\label{eq:jordanblock}
\end{equation}
where $D$ is the matrix with unit entries on the first superdiagonal, i.e., $D_{nn'}=\delta_{n,n'-1}$. An important feature of these matrices is that they are nilpotent with index $n_l$, hence, $D^{n_l}=0$ but $D^{n_l-1} \neq 0$. For an isolated eigenvalue, we simply have $D=0$, so that this term can be dropped from the decomposition. If this applies to all eigenvalues,  this aligns with Eq.~\eqref{eq:decnonh1} after applying the similarity transformation \eqref{eq:jordandecomp}.

In a given basis of eigenvectors and covectors, the matrix elements of the transformation matrix $T$ can be determined as $T_{m,ln}=\langle m |J^{(l)}_n\rangle$, where the second multi-index refers to the described block structure of $J$, with $l$ running over the distinct eigenvalues and $n$ running from 1 to $n_l$.
In turn, given a Jordan decomposition \eqref{eq:jordandecomp}, a right Jordan chain is obtained from the columns of the matrix $T$, while the corresponding biorthogonal left Jordan chain is then obtained from the rows of the matrix $T^{-1}$.

As before, this construction can be interpreted as a generalized spectral decomposition,
\begin{equation}
    \op{H}=\sum_l(E_l\hat P_l+\hat N_l),
    \label{eq:decnonh2}
\end{equation}
where 
\begin{equation}\label{eq:projectorPl}
\hat P_l=\sum_{n=1}^{n_l}
|J^{(l)}_n\rangle\langle \tilde J^{(l)}_n|
\end{equation}
arises from the identity matrix in the Jordan block \eqref{eq:jordanblock}, while 
\begin{equation}
\hat N_l=\sum_{n=1}^{n_l-1}
|J^{(l)}_n\rangle\langle \tilde J^{(l)}_{n+1}|
\label{eq:nilpotent}
\end{equation}
arises from the matrix $D$ in this block.

While the construction of the biorthogonal Jordan chains allows for some freedom, which is also reflected in the detailed form of the transformation matrix $T$, the operators $\hat P_l$ and $\hat N_l$ are unique. From their construction, the operators $\hat P_l$ are oblique projectors onto the (generalized) eigenspace associated with the specified eigenvalue $E_l$ that continue to fulfill Eq.~\eqref{eq:projcond}, but now are of finite rank $n_l$. Furthermore, the operators $\hat N_l$ are nilpotent with index $n_l$, i.e., $\hat N_l^{n_l}=0$ but $\hat N_l^{n_l-1}\neq 0$. Additionally, these operators obey the important relation
\begin{equation}
\hat N_l=(\hat H-E_l\hat I)\hat P_l.
\label{eq:nfromh}
\end{equation}
Tuning a non-Hermitian system to an EP therefore drastically changes the mathematical nature of the eigenvalue problem. 

\subsection{Measures of mode nonorthogonality for isolated resonances\label{sec:measures}}
The spectral decompositions described above imply a strong link between the spectral features of non-Hermitian systems and the geometric structure of the states associated with them. To prepare the discussion of these features at EPs, we survey common theoretical mode-nonorthogonality measures that illuminate different aspects of this link for spectrally isolated resonances, and detail how they are closely related to each other. The detailed physical contexts for these quantities are provided in later sections, where we generalize these notions to apply to systems at EPs.

For a system with a nondegenerate spectrum that can be diagonalized as given in Eq.~\eqref{eq:diagnonh}, it is useful to introduce the mode overlap matrix~\cite{CM98,Wiersig19} $O_{kl}=(V^\dagger V)^{-1}_{kl}(V^\dagger V)_{lk} $, which encodes the departure of the eigensystem from an orthonormal system. The diagonal elements
\begin{equation}
O_{ll}=K_l=\langle L_l|L_l\rangle\langle R_l|R_l\rangle
\label{eq:petermann}
\end{equation}
(subject to biorthogonal normalization
$\langle L_l|R_l\rangle=1$)
are known as Petermann factors~\cite{Petermann79,Siegman86,SFP00,Schomerus09,CS12,PNS14}. 
The Petermann factor equals unity, $K_l=1$, when $\langle L_l|^\dagger\propto |R_l\rangle$ 
(i.e., equal up to normalization),
which is automatically fulfilled in Hermitian systems, while generally $K_l\geq 1$.
Geometrically, this factor therefore quantifies the degree of nonorthogonality of the eigenstates associated with a specific isolated eigenvalue.
This definition also extends to the isolated resonance in systems where other eigenvalues may participate in EPs.
Indeed, avoiding the requirement of biorthogonal normalization, the Petermann factor can be written as
\begin{equation}
    K_l=\frac{\langle L_l|L_l\rangle\langle R_l|R_l\rangle}{|\langle L_l|R_l\rangle|^2}. 
\label{eq:petermann2}    
\end{equation}

The Petermann factor appeared first in studies of open-cavity lasers~\cite{Petermann79}, where it characterizes excess noise due to cross-talk of the overlapping cavity eigenmodes. The originally studied systems were reciprocal, so that the effective non-Hermitian  Hamiltonian obeys $H=H^T$ in a suitable basis~$|m\rangle$. For such reciprocal systems, one can identify the right and left eigenvectors as $\langle L_l|m\rangle=\langle m|R_l\rangle$, so that the Petermann factor takes the form
\begin{equation}
K_l=
\frac{(\sum_m |\langle m|R_l\rangle|^2)^2}{|\sum_m \langle m|R_l\rangle^2|^2}\quad
\mbox{(reciprocal systems).}
\end{equation}
In this form, the Petermann factor equals unity when the components of the eigenstate in the reciprocal basis can all be taken as real. This links it to a measure of wavefunction complexity, the phase rigidity
\begin{equation}
r_l=
\frac{|\sum_m \langle m|R_l\rangle^2|}{\sum_m |\langle m|R_l\rangle|^2}\quad
\mbox{(reciprocal systems),}
\end{equation}
so that $K_l=r_l^{-2}$. This complexity measure characterizes, e.g., the imbalance of counter-propagating wavefunction components  in a plane-wave decomposition, as frequently encountered in the decay out of a scattering region~\cite{LBB97}. At the same time, the phase rigidity coincides exactly with the mathematical definition of the eigenvalue condition number, which characterizes the sensitivity of the eigenvalue problem against generic perturbations.
Retracing these steps, the phase rigidity is naturally generalized to non-reciprocal systems by setting $r_l=K_l^{-1/2}$, see, e.g., Ref.~\cite{Wiersig23}. This abandons its original interpretation as a measure of complex eigenstate components, but maintains the exact agreement with the mathematical definition of the eigenvalue condition number in this case.

Even though formulated from different perspectives, the different mode-nonorthogonality measures
are therefore strictly related to each other.
Hence, for isolated resonances, physical considerations of mode crosstalk, wavefunction imbalances, and eigenvalue sensitivity  all relate to
a very specific and universal aspects of mode nonorthogonality.
In the following sections, we will explore how a richer and more complex picture emerges at EPs.

\subsection{Mode nonorthogonality near exceptional points\label{sec:modeEP}}
Even though the  mode-nonorthogonality measures of the previous section have been formulated for isolated resonances, they also give insights into the drastic modifications that occur as a system approaches an EP.
The coalescence of eigenvectors at EPs implies a breakdown of the biorthogonality condition \eqref{eq:biorth}, thus yielding a self-orthogonal state $\langle L_l|R_l\rangle=0$. Therefore, the Petermann factor \eqref{eq:petermann2} diverges at these EPs, while the phase rigidity and the condition number vanish at these points.

A careful perturbative analysis~\cite{Wiersig23,HS24} reveals that the asymptotic behavior of the mode-nonorthogonality measures near EPs is tied to the algebraic structure of the system at the EP. Formulated in terms of the Petermann factor, one then finds that near an EP of order $n_l$
\begin{equation}
K_{l'}\sim \frac{\xi_l^2}{n_l^2|E'_{l'}-E_{l}|^{2n_l-2}},
\label{eq:kdivergence}
\end{equation}
where $E_{l}$ is the degenerate eigenvalue in the system with the EP, and
$E'_{l'}$ any one of the $n_l$ quasi-degenerate but now  isolated eigenvalues in a generically perturbed system. The spectral response strength 
\begin{equation}
  \xi_l=||\hat N_l^{n_l-1}||_2  
  \label{eq:responsestrength}
\end{equation}
arises from the nilpotent part of the spectral decomposition at the EP, with the spectral norm $||\hat A||_2=\max_{|\psi\rangle\neq 0}\, ||\hat A|\psi\rangle||_2/||\psi||_2$. The same numbers $n_l$ and $\xi_l$ apply equally to all quasi-degenerate eigenvalues  in the perturbed system, hence, are independent of the index $l'$ distinguishing these eigenvalues.

Therefore, there is a close relation between the geometry of eigenvectors near an EP, and the Jordan decomposition at the EP. A remarkable aspect of this relation is the fact that the Jordan decomposition itself is mathematically ill-conditioned against perturbations that lift the EP degeneracy. As soon as the degeneracy is lifted, the nature of the spectral decomposition changes drastically. The same complications arise in practical applications, such as in the numerical analysis of specific systems. In contrast, given the relation \eqref{eq:kdivergence}, the spectral strength itself is a well-behaved quantity that can also be evaluated slightly away from the EP. The spectral strength can indeed be interpreted as a well-behaved algebraic quantity \cite{BS25} --- and this well-behaved nature is also in keeping with its direct physical significance for the response of the system to perturbations and noise.

This naturally raises the question which features of the system  determine the magnitude of the spectral strength, and in particular how this relates to the mathematical formulation and physical role of mode nonorthogonality at the EP itself.
Mathematically, this geometric structure is encoded in the transformation to the Jordan form, which generally is nonunitary. Just as for the case of isolated resonances, the similarity transformation $T$ hides important physical aspects of the system, namely all those related to the nonorthogonality of the eigenvectors and Jordan chains.   

The preceding two paragraphs outline the issues that we tackle in the present work. 
To resolve them, we will adopt the perspective of spectral response theory,  and show, in the next section, how this leads to a natural generalization of the Petermann factor that reliably characterizes the mathematical nature and physical role of mode nonorthogonality both at and near EPs, as well as the connection  between these two situations.

\section{Response perspective}
\label{sec:response}
\subsection{Setting the scene}
Mathematically, the surveyed mode-nonorthogonality measures for isolated resonances characterize the intrinsic geometric relation of a given pair of right and left eigenvectors. At the same time, these eigenvectors are subject to the biorthogonality condition  
\eqref{eq:biorth}, hence, also capture the extrinsic interplay with eigenvectors of other eigenvalues.
We can make this manifest by  expressing the Petermann factor \eqref{eq:petermann2} of a given isolated eigenvalue as 
\begin{equation}
K_l=\mbox{tr}\, \op{P}_l^\dagger \op{P}_l, \quad\mbox{(isolated eigenvalue)}
\label{eq:kfromp}
\end{equation}
hence, directly in terms the uniquely defined oblique projection operator on the associated eigenspace, and noting that the collection of these projection operators is further constrained by Eq.~\eqref{eq:projcond}.

Physically, the Petermann factor appears naturally in this form when we consider the generic  response of the system, 
and then captures deviations from  conventional Breit-Wigner resonance relations. 
We formulate this in terms of the spectrally resolved response power 
\begin{equation}
\mathcal{P}(E)=\mathrm{tr}\, \op{G}^\dagger(E)\op{G}(E),
\label{eq:power}
\end{equation}
where $\op{G}(E)=(E\op{I}-\op{H})^{-1}$ is the Green's function of the effective Hamiltonian.
For a non-degenerate  system, the spectral decomposition \eqref{eq:decnonh1} gives 
\begin{equation}
\op{G}(E)=\sum_l\frac{1}{E-E_l}\op{P}_l.
\label{eq:decnong1}
\end{equation}
For energies $E$ close to an eigenvalue $E_l$, one then obtains 
\begin{equation}
\mathcal{P}(E)\sim \frac{K_l}{|E-E_l|^2},
\quad\mbox{(isolated eigenvalue)}
\label{eq:pisolated}
\end{equation}
where $\sim$ denotes exact asymptotic equality for energies where the contributions from all other eigenvalues can be neglected, and $K_l$ is the Petermann factor given by Eq.~(\ref{eq:petermann2}) or \eqref{eq:kfromp}.

In Breit-Wigner resonance theory, the effects of mode nonorthogonality are neglected, which corresponds to setting $K_l=1$. The Petermann factor therefore signifies an enhanced response due to the nonorthogonality of the eigenstates when compared to a system with orthogonal eigenstates---such as given by the diagonalized Hamiltonian.  
This difference is tightly linked to the fact that the transformation \eqref{eq:diagnonh} diagonalizing the Hamiltonian is not unitary.

\subsection{Petermann factor of exceptional points}
Near an isolated eigenvalue, the asymptotic form \eqref{eq:pisolated} holds even when some of the other eigenvalues partake in EPs, as can be read off the more general spectral decomposition in Eq.~\eqref{eq:decnonh2}.
The corresponding expansion of the Green's function then reads
\cite{Wiersig23b} 
\begin{equation}\label{eq:GKato}
	\op{\GF}(\ev) = \sum_\ind\left[\frac{\projector_\ind}{\ev-\ev_\ind} + \sum_{k=2}^{\order_\ind} \frac{\op{\nilpotent}_\ind^{k-1}}{(\ev-\ev_\ind)^k}\right] \ .
\end{equation}

Let us now consider the response near a degenerate eigenvalue $E_l$, corresponding to an EP of order $\order_\ind$, with oblique projector $ \projector_\ind$ and nilpotent matrix $\nilpotent_\ind$. For energies $\ev\approx E_l$ the dominant contribution to the Green's function of the EP is in this case~\cite{Wiersig22}
\begin{equation}\label{eq:domicontri}
	\op{\GF}(E) \sim \frac{\op{\nilpotent}^{n_l-1}_l}{(\ev-E_l)^{n_l}} 
\end{equation}
with rank-1 matrix $\nilpotent^{n_l-1}_l$. 
The qualitative change of the energy dependence amounts to a modified resonance lineshape,  marking a complete departure from Breit-Wigner resonance theory that has been encountered both in specific theoretical and experimental settings~\cite{YSS11,PZM17,HKC22,SZM22}.
By its mathematical definition \eqref{eq:responsestrength}, the spectral response strength $\xi_l$ then characterizes the maximal response of the system within this modified resonance behavior.

We now connect this behavior to the mode nonorthogonality at the EP. For this, we first note that 
for rank-1 matrices such as $\nilpotent^{n_l-1}_l$, the spectral norm gives the same result as the Frobenius norm $\normFro{A}=(\mbox{tr}\,A^\dagger A)^{1/2}$. Hence,
\begin{equation}
\xi_l^2=\mbox{tr}[(\hat N_l^{n_l-1})^\dagger \hat N_l^{n_l-1}] .
\label{eq:kgeneralisedold}
\end{equation}
With Eq.~\eqref{eq:nfromh} and careful utilization of the spectral decomposition  \eqref{eq:decnonh2}, this factor can further be put into the form
\begin{equation}
\xi_l^2
=\mbox{tr}\,\hat P_l^\dagger[(\hat H^\dagger-E_l^*\hat I)]^{n_l-1} [(\hat H-E_l\hat I)]^{n_l-1}\hat P_l,
\label{eq:kgeneralised3}
\end{equation}
which reveals this to be related to an oblique projection of the system.
We can compare Eq.~(\ref{eq:kgeneralised3}) to
\begin{align}
&\left(\xi_l^{(R)}\right)^2=
\label{eq:kgeneralised3R}
\\
&\,\,\quad\mbox{tr}\,[\hat P_l^{(R)}(\hat H^\dagger-E_l^*\hat I)\hat P_l^{(R)}]^{n_l-1} [\hat P_l^{(R)}(\hat H-E_l\hat I)\hat P_l^{(R)}]^{n_l-1}
\nonumber
\end{align}
and 
\begin{align}
&\left(\xi_l^{(L)}\right)^2=
\label{eq:kgeneralised3L}
\\
&\,\,\quad\mbox{tr}\,[\hat P_l^{(L)}(\hat H^\dagger-E_l^*\hat I)\hat P_l^{(L)}]^{n_l-1} [\hat P_l^{(L)}(\hat H-E_l\hat I)\hat P_l^{(L)}]^{n_l-1}  
\nonumber
\end{align}
where $\hat P_l^{(R)}$ and $\hat P_l^{(L)}$ are the projectors on the spaces spanned by the right and left Jordan chains.
These projectors can be written explicitly as
\begin{align}
\hat P_l^{(R)}&=\sum_{km}A_{km}
|J^{(l)}_k\rangle\langle J^{(l)}_m|
,
\label{eq:explicitPR}
\\
\hat P_l^{(L)}&=\sum_{km}\tilde A_{km}
|\tilde J^{(l)}_k\rangle\langle \tilde J^{(l)}_m|,
\label{eq:explicitPL}
\end{align}
where $A=B^{-1}$, $\tilde A =\tilde B^{-1}$ are obtained by inverting Gram matrices with elements
$    B_{km}=\langle J^{(l)}_k|J^{(l)}_m\rangle$,
    $ \tilde B_{km}=\langle \tilde J^{(l)}_k|\tilde J^{(l)}_m\rangle
    $. Equations 
   \eqref{eq:kgeneralised3R} and \eqref{eq:kgeneralised3L} then represent orthogonal truncations of the nilpotent part $\hat N_l$ to the right or left generalized eigensubspaces.
These truncations would be valid if these subspaces are orthogonal to the subspaces of the remaining vectors. The difference is due to mode-nonorthogonality.

The physical meaning of these truncations is revealed by considering the response of the system.
For energies $E\sim E_l$, the spectrally resolved response power \eqref{eq:power} is given by 
\begin{equation}
\mathcal{P}(E)\sim \frac{\xi_l^2}{|E-E_l|^{2n_l}}
\quad\mbox{(system at EP-$n_l$)}.
\end{equation}
Hence, we interpret the pair of values
\begin{equation}
K_l^{(R,L)} = \left(\frac{\xi_l}{\xi_l^{(R,L)}}\right)^2
\label{eq:kgeneralised}
\end{equation}
as a generalized \textit{EP Petermann factor} that captures the enhanced response of the system at an EP of order $n_l$ compared to the orthogonal truncated reference systems. 
The two alternative truncations have a concrete physical interpretation, as the right subspace determines the profile of the detected response, while the left subspace determines the coupling to the source that excites the system \cite{HS20}.
These interpretations then carry over to the Petermann factors $K_l^{(R)}$ and $K_l^{(L)}$. 

As shown in Appendix~\ref{app:kgeom}, the EP Petermann factors \eqref{eq:kgeneralised} can further be expressed in purely geometric terms as
\begin{align}
K_l^{(R)}=\frac{\langle L_l|L_l\rangle}{\langle L_l|\hat P_l^{(R)}|L_l\rangle}
=\frac{1}{\mbox{tr}\,\hat Q_l^{(L)}\hat P_l^{(R)}}
,
\label{eq:explicitR}
\\
K_l^{(L)}=\frac{\langle R_l|R_l\rangle}{\langle R_l|\hat P_l^{(L)}|R_l\rangle}
=\frac{1}{\mbox{tr}\,\hat Q_l^{(R)}\hat P_l^{(L)}},
\label{eq:explicitL}
\end{align}
where 
$\hat Q_l^{(R)}=\frac{|R\rangle\langle R|}{\langle R|R\rangle}$ and
$\hat Q_l^{(L)}=\frac{|L\rangle\langle L|}{\langle L|L\rangle}$
are the orthogonal rank-1 projectors onto the right and left eigenvector. For isolated resonances with $n_l=1$, where $\hat Q_l^{(R)}=\hat P_l^{(R)}$
and $\hat Q_l^{(L)}=\hat P_l^{(L)}$,
the expressions in Eqs.~(\ref{eq:explicitR})-(\ref{eq:explicitL}) reduce to the conventional Petermann factor in Eq.~\eqref{eq:kfromp}.
We see that the EP Petermann factor obeys
$K_l^{(R,L)}\geq 1$. Furthermore, the limiting case $K_l^{(R)}= 1$ occurs when the left eigenvector lies completely in the space spanned by the right generalized eigenvectors, which by the biorthogonality condition 
\eqref{eq:jordanbiorth} implies that it is orthogonal to the left generalized eigenvectors associated with all remaining eigenvalues; and the same condition applies mutatis mutandi to  $K_l^{(L)}= 1$. These observations  underline the geometric interpretation of this quantity.

Furthermore, these physical and geometric considerations suggest to combine the two Petermann factors via their geometric mean into a single overall Petermann factor
\begin{equation}
K_l\equiv \sqrt{K_l^{(R)}K_l^{(L)}}=\frac{\xi_l^2}{\xi_l^{(R)}\xi_l^{(L)}}.
\label{eq:koverallphysical}
\end{equation}
Expressed in terms of the geometric data, this takes the instructive form
\begin{equation}
K_l^2=\frac{\langle R_l|R_l\rangle\langle L_l|L_l\rangle}{\langle R_l|\hat P_l^{(L)}|R_l\rangle \langle L_l|\hat P_l^{(R)}|L_l\rangle}.
\label{eq:koverallgeometrical}
\end{equation}
This combined Petermann factor again reduces to the conventional Petermann factor for $n=1$. We see that still $K_l\geq 1$, while $K_l=1$ now gives a sharper orthogonality criterion combining the criteria for $K_l^{(R)}$ and $K_l^{(L)}$. We will subsequently see that each of these three numbers plays a useful role in characterizing the mode-nonorthogonality at an EP.

To further illuminate this generalization, we evaluate the spectral response strength based on the detailed construction of the nilpotent operator $\hat N_l$ in the original basis, as given by Eq.~\eqref{eq:nilpotent}. This delivers 
\begin{equation}
\hat N_l^{n_l-1}=|J^{(l)}_1\rangle\langle \tilde J^{(l)}_{n_l}|=|R_l\rangle\langle L_l|,
\label{eq:nnl1}
\end{equation}
where we utilized the Jordan biorthogonality condition \eqref{eq:jordanbiorth}, and the anchoring of the Jordan chains at the right and left eigenvectors. 
Introducing this into Eq.~\eqref{eq:kgeneralisedold}, we obtain the remarkable result that the spectral response strength 
\begin{equation}
\rca_l^2 = \langle R_l|R_l\rangle\langle L_l|L_l\rangle
\label{eq:kgeneralised2}
\end{equation}
formally replicates the expression \eqref{eq:petermann} of the conventional Petermann factor. However, crucially, instead of fulfilling the conventional biorthogonality conditions \eqref{eq:biorth}, including $\langle L_l|R_l\rangle=1$, these vectors are now subject to the Jordan biorthogonality relations \eqref{eq:jordanbiorth}, which instead include the selforthogonality condition $\langle L_l|R_l\rangle=0$. 
Importantly, despite its suggestive form \eqref{eq:kgeneralised2} the resulting spectral response strength is dimensionful, hence, does not serve as a purely geometric measure of mode nonorthogonality. 
Indeed, the dimensionful analogues of the spectral response strength are the quantities given in \eqref{eq:kgeneralised3R} and 
\eqref{eq:kgeneralised3L}, which we can now write as (see Appendix~\ref{app:kgeom})
\begin{align}
\left(\xi_l^{(R)}\right)^2
&=\langle R_l|R_l\rangle\langle L_l|
\hat P_l^{(R)}
|L_l\rangle
,
\label{eq:xiRcompact}
\\
\left(\xi_l^{(L)}\right)^2
&=
\langle R_l|
\hat P_l^{(L)}
|R_l\rangle\langle L_l|L_l\rangle
\label{eq:xiLcompact}
.
\end{align}
As we will see in their concrete evaluation, these dimensionful quantities are naturally interpreted as \textit{internal response strengths}.
In contrast, the generalized Petermann factor \eqref{eq:kgeneralised} is dimensionless, as is required for a purely geometric quantity.

We therefore conclude that just as for isolated resonances, the physical response of systems at an EP is enhanced by mode nonorthogonality,
\begin{equation}
\mathcal{P}(E)\sim \frac{\left(\xi_l^{(R,L)}\right)^2K_l^{(R,L)}}{|E-E_l|^{2n_l}},
\quad\mbox{(EP)}
\label{eq:pep}
\end{equation}
which can be quantified by a pair of generalized Petermann factors $K_l^{(R,L)}$ that are directly related to the spectral response strength.
Concretely, these Petermann factors can then be isolated by comparison of total response power in Eq.~\eqref{eq:pep} with the conditioned response powers
\begin{align}
\mathcal{P}_R(E)=\mathrm{tr}\, \op{G}^\dagger(E)\op{G}(E) \hat P^{(R)}
\sim \frac{\left(\xi_l^{(R)}\right)^2}{|E-E_l|^{2n_l}},
\label{eq:pepr}
\\
\mathcal{P}_L(E)=\mathrm{tr}\, \op{G}^\dagger(E)\hat P^{(L)}\op{G}(E) 
\sim \frac{\left(\xi_l^{(L)}\right)^2}{|E-E_l|^{2n_l}},
\label{eq:pepl}
\end{align}
corresponding to a source covering the right eigenspace and a detector covering the left eigenspace, respectively.
The geometric content of this enhancement is emphasized by its  explicit form  \eqref{eq:explicitR}, \eqref{eq:explicitL}, while 
Eq.~\eqref{eq:kgeneralised} affords the stated physical interpretation. Finally, these EP Petermann factors can be naturally combined into an overall EP Petermann factor $K_l$, expressed physically in terms of enhanced response in Eq.~\eqref{eq:koverallphysical} and geometrically in terms of mode-nonorthogonality in Eq.~\eqref{eq:koverallgeometrical}.

Note that our EP Petermann factor is very different from the generalized Petermann factor in Ref.~\cite{ZZC24}. The latter is an ad hoc order parameter measuring of global non-unitarity of a non-Hermitian system, formulated purely in terms of the right eigenvectors and constructed to not diverge at EPs.

\subsection{Consistency with the behavior near exceptional points}
While the generalized EP Petermann factor is finite, it is only strictly defined exactly at the EP. On the other hand, the conventional Petermann factor diverges when an isolated eigenvalue approaches an EP.
Nevertheless, the spectral response detailed in the previous subsection provides a direct link between the mode nonorthogonality at EPs and the behavior near EPs, embodied by Eq.~\eqref{eq:kdivergence}. To conclude our general considerations, we establish the consistency of this link. 

For this, we note that under the generic perturbative conditions where Eq.~\eqref{eq:kdivergence} applies, the $n_l$ quasi-degenerate eigenvalues  $E'_{l'}$ are spread out equidistantly around a circle in the complex plane, centered at the position $E_{l}$ of the EP in the original, degenerate system; see, e.g., Ref.~\cite{KGK23}. Providing these eigenvalues with the indices $l'=1,2,3,\ldots,n_l$, this can be formalized by approximating $E'_{l'}\approx E_{l}+w^{l'}\Delta $, where $w=\exp(2\pi i/n_l)$ is a root of unity, and $\Delta$ is a complex parameter whose phase provides freedom of the location of the eigenvalues on the circle, while $|\Delta|$ is assumed to be small enough so that other eigenvalues remain well separated. 

These conditions provide an energy window in which $|\Delta|\ll |E-E_{l}|,|E-E'_{l'}|\ll|E-E_{m}|,|E-E'_{m'}|$, with the index $l'$ again restricted to the quasi-degenerate eigenvalues,  and the indices $m$ and $m'$ referring to the other eigenvalues of the unperturbed and perturbed system (which likewise remain close to each other). 
Physically, this examines the collective response of the quasi-degenerate cloud of eigenvalues at an energetic distance much larger than the energy splitting in the cloud, while other resonances can still be neglected.

Under these conditions, we can approximate the collective resonant contribution to the Green's function as 
\begin{align}
\op{G}(E)\sim\sum_{l'=1}^{n_l}\frac{1}{E-E'_{l'}}\op{P'}_{l'},
\label{eq:resclosetoep}
\end{align}
where the projectors $\op{P'}_{l'}$ become related to each other as the right and left eigenvectors of all quasidegenerate eigenvalues approach each other at the EP.

The relation between these projectors concerns their structure, magnitude and their complex phase. For the structure, we observe that all projectors become aligned according to 
$\op{P'}_{l'}\propto \op{N}_l^{n_l-1}$, cf.~Eqs.~\eqref{eq:decnonh1} and \eqref{eq:nnl1}. As emphasized before, the  eigenvectors in both expressions are subject to different biorthogonality conditions [see Eq.~\eqref{eq:jordanbiorth} for the Jordan chain].
For the magnitude of the ensuing proportionality constant,
we note that according to Eq.~\eqref{eq:kdivergence},
all eigenvalues have asymptotically identical Petermann factors $K_{l'}=\mbox{tr}\, \op{P'}_{l'}^\dagger \op{P'}_{l'}\approx|\xi_l|^2/n_l^2|\Delta|^{2n_l-2}$. 
To furthermore capture the complex phase relations between all terms,  we introduce the residues 
\begin{equation}
s_{l'}\sim\frac{\xi_l}{n_l(E'_{l'}-E_l)^{n_l-1}}
\sim \frac{\xi_l}{n_l}\frac{1}{\Delta^{n_l-1}w^{-l'}},
\end{equation}
where we used $w^{n_l-l'}=w^{-l'}$ as $w^{n_l}=1$.
These residues fulfill $|s_{l'}|^{2}=K_{l'}$ in agreement with
Eq.~\eqref{eq:kdivergence}, hence $|s_{l'}^{-1}|=|r_{l'}|$ in terms of the phase rigidity, illustrating a case where the latter quantity is usefully generalized to include a complex phase.
In the present case, these complex phases are then determined by the fractional Berry phase of the EP~\footnote{Within our the response framework, this Berry phase can be determined form the invariance of the resonant part \eqref{eq:resclosetoep} of the Green's function  when one encircles the EP in parameter space, after which the eigenvalues are cyclically interchanged.}.

Utilizing the identity for an arbitrary complex constant $a$ (see Appendix~\ref{app:b})
\begin{equation}\sum_{l'=1}^{n_l}\frac{1}{aw^{l'}-1}=\frac{n_l}{a^{n_l}-1} \ ,
\label{eq:identity}
\end{equation}
the collective resonant contribution to the spectrally resolved response power  \eqref{eq:power} becomes
\begin{align}
\mathcal{P}(E)\sim&\left|\frac{\xi_l}{n_l}\sum_{l'=1}^{n_l}\frac{1}{\Delta^{n_l-1}w^{-l'}}\frac{1}{E-E_{l}-\Delta w^{l'}}\right|^2 
\nonumber\\
&=\left|\frac{\xi_l}{(E-E_{l})^{n_l}-\Delta^{n_l}}\right|^2 \quad\mbox{(system near EP-$n_l$)}.
\end{align}
For $\Delta\to 0$, where one approaches the EP, this indeed reduces to Eq.~\eqref{eq:pisolated}.

Therefore, despite the singular mathematical change in the algebraic description of systems at and near EPs, and the ensuing  qualitative departure from the Lorentzian resonance lineshapes of Breit-Wigner theory, the detailed geometric features of mode nonorthogonality play out to provide a physically consistent response of the system. 

\section{Quantum-limited excess noise}
\label{eq:noise}
\subsection{Setting the scene}
In its explicit relation to mode non-orthogonality, the Petermann factor first appeared in the theory of quantum-limited emission of  open-cavity amplifiers and lasers. In these systems, the quantum limit is set by spontaneous emission processes that supply photons with an undetermined phase relation with respect to coherent photons from stimulated emission. The physical manifestation is a lower bound of the linewidth $\Delta\omega$.
This bound can be evaluated in an idealized setting where other factors, such as inhomogeneous broadening, incomplete population inversion in the medium, and complex frequency dispersion (the linewidth enhancement Henry alpha factor) are neglected. 
Universal statements can then be made in the regime of amplified spontaneous emission (ASE) just below the laser threshold, where sharp resonances already appear but nonlinearities can still be neglected, and conversely sufficiently far above the laser threshold (but still in the regime of single-mode lasing), where nonlinear feedback eliminates the intensity fluctuations due to the noisy photons, which effectively halves the linewidth. 

Treating the described cases in conventional Breit-Wigner theory (hence, neglecting the role of mode non-orthogonality), the spectrally resolved output intensity
\begin{equation}
    I(\omega)=\frac{I}{2\pi}\frac{\Delta\omega}{(\omega-\Omega_0)^2+\Delta\omega^2/2},
\end{equation}
arising from an isolated cold-cavity mode with complex 
frequency
 $\omega_0=\Omega_0-i\Gamma_0/2$
is a Lorentzian of a pump-strength dependent total intensity $I=\int I(\omega) d\omega$ and linewidth $\Delta\omega$, which are related by 
\begin{align}
\Delta \omega=\frac{\Gamma_0^2}{2I}\equiv \Delta \omega_{ST} \quad\mbox{(lasing case)}
\label{eq:deltast}
\end{align}
and 
\begin{align}
\Delta \omega=\frac{\Gamma_0^2}{I}=2\Delta \omega_{ST} \quad\mbox{(ASE case)},
\label{eq:deltast2}
\end{align}
where the  expression $\Delta \omega_{ST}$ for the laser linewidth is known as the Schawlow-Townes formula.

Mode-nonorthogonality can be accounted for by determining the dynamical response of the system to the quantum noise using  the Green's function from the previous section. The appropriate noise strength can be established within the input-output formalism, which exploits the bosonic commutation relations to express the quantum noise in terms of the dissipative losses (a version of the fluctuation theorem) \cite{SFP00}. 
In the regime of ASE just below the laser threshold, one then obtains, as a key intermediate step, the spectrally resolved output intensity
\begin{equation}
I(\omega)\approx\frac{1}{2\pi}\mathrm{tr}\,[\hat\Gamma \hat  G(\omega-i\gamma/2)]^\dagger
\hat\Gamma \hat G(\omega-i\gamma/2),
\label{eq:asestart}
\end{equation}
with once again $\op{G}(E)=(E\op{I}-\op{H})^{-1}$, where $\hat H$  now specifically
represents the effective non-Hermitian Hamiltonian in the open cold cavity, $i\hat \Gamma=\hat H-\hat H^\dagger$ the decay due to the coupling to the environment, i.e., radiative losses, and $\gamma$ the rate of spatially uniform gain supplied by an ideal amplifying medium. The latter rate is tuned so that the ASE emission resonance is shifted close to the real axis, which results in a well-defined ASE signal.

The Petermann factor then appears when one evaluates this expression for a spectrally isolated emission resonance with frequency $\omega_0=\Omega_0-i\Gamma_0/2$ and biorthogonal right and left eigenstates $|R_0\rangle$, $\langle L_0|$. Neglecting the contributions of all other resonances, we can approximate
$G(\omega)\approx |R_0\rangle(\omega-\omega_0)^{-1}\langle L_0 |$
with biorthogonality condition $\langle L_0|R_0\rangle = 1$ and then  obtain
\begin{align}
I(\omega)&\approx \frac{1}{2\pi}
\langle L_0|\hat\Gamma|L_0\rangle\frac{1}{\omega-\Omega_0+i(\gamma-\Gamma_0)/2}
\nonumber\\
&\times
\langle R_0|\hat\Gamma|R_0\rangle\frac{1}{\omega-\Omega_0-i(\gamma-\Gamma_0)/2}.
\label{eq:ase1}
\end{align}
With the sum rule 
\begin{align}\label{eq:sumlaser}
  \Gamma_0&=\frac{\langle L_0|\hat\Gamma|L_0\rangle}{\langle L_0|L_0\rangle}=\frac{\langle R_0|\hat\Gamma|R_0\rangle}{\langle R_0|R_0\rangle}
\end{align}
relating the decay rate to the coupling strength, we can write
\begin{align}\label{eq:sumlaser2}
  \langle R_0|\hat\Gamma|R_0\rangle\langle L_0|\hat\Gamma|L_0\rangle=K_0
  \Gamma_0^2,
\end{align}
where $K_0$ is the Petermann factor as introduced in Eq.~\eqref{eq:petermann}.
Therefore, the spectrally resolved ASE emission intensity \begin{align}
&I(\omega)=
\frac{1}{2\pi}\frac{K_0\Gamma_0^2}{(\omega-\Omega_0)^2+(\gamma-\Gamma_0)^2/4},
\end{align}
is a Lorentzian of width $\Delta \omega=\gamma-\Gamma_0$ and total intensity $I=K_0{\Gamma_0^2}/{\Delta\omega}$, so that now
\begin{align}
\Delta \omega=2K\Delta \omega_{ST} \quad\mbox{(ASE case)},
\label{eq:deltast3}
\end{align}
while sufficiently far above the laser threshold feedback reduces this again by a factor of 2, to
\begin{align}
\Delta \omega=K \Delta \omega_{ST} \quad\mbox{(lasing case)}.
\label{eq:deltast4}
\end{align}
We see that the Petermann factor arises from the matrix elements  \eqref{eq:sumlaser} 
of the anti-Hermitian part
$i\hat \Gamma$ of the effective cold-cavity Hamiltonian, which characterizes the openness of the system. On the one hand, this reflects the nature of the physical problem, where the strength of the quantum noise depends on the coupling to external vacuum fluctuations. From this physical perspective, the excess noise can then be attributed to crosstalk with nonlasing modes that overlap with the lasing mode.
We note that this quantum limit applies generally also to disordered or wave-chaotic systems \cite{SFP00,Schomerus09}, while  experimental considerations to approach this limit can be found in Refs.~\cite{Cheng1996,Eijkelenborg1996}.

\subsection{Generalization to exceptional points}
We now evaluate the quantum-limited excess noise near an EP of order $n_l$, where we again denote the
frequency $\omega_l=\Omega_l-i\Gamma_l/2$, and the right and left eigenstates $|R_l\rangle$, $\langle L_l|$.
The starting point is the general expression in Eq.~\eqref{eq:asestart}, into which we now introduce the approximation
\eqref{eq:domicontri}. Utilizing Eqs.~(\ref{eq:nnl1}), (\ref{eq:kgeneralised2}), and \eqref{eq:sumlaser}, we obtain the resonant ASE emission spectrum
\begin{equation}
    I(\omega)=\frac{1}{2\pi}\frac{\Gamma_l^2\xi^2}{[(\omega-\Omega_0)^2+(\gamma-\Gamma_l)^2/4]^{n_l}},
    \label{eq:Iomegaxi}
\end{equation}
which is of the form a super-Lorentzian with width $\Delta\omega=(\gamma-\Gamma_l)$ and total intensity
\begin{equation}
I=\frac{[2(n_l-1)]!}{[(n_l-1)!]^2}\frac{\Gamma_l^2\xi^2}{\Delta\omega^{2 n_l-1}}
.
    \label{eq:Ixi}
\end{equation}
It is emphasized that neither the intensity in Eq.~(\ref{eq:Iomegaxi}) nor  the width in Eq.~(\ref{eq:Ixi}) diverges to infinity at the EP. This recovers the behavior from specific model scenarios \cite{YSS11}, generalizes this to EPs or any order, and reexpresses this in terms of the general quantities studied in the present work. 
The key point is that the excess noise intensity at the EP depends on the spectral strength $\xi_l^2$. 
Therefore, the EP Petermann factors developed in the previous section directly carry over to this setting.  
In particular, these EP Petermann factors can in principle be isolated by suitable design of the source (now given by the gain medium) or detector, 
following the same prescription as in Eqs.~\eqref{eq:pepr} and \eqref{eq:pepl}.
This extends the general picture that quantum noise is enhanced by mode non-orthogonality to systems operating at EPs. In particular, just as in the original setting of isolated resonances, the EP Petermann factors naturally quantify the noise crosstalk with other modes that overlap with the lasing mode; however, by employing the notions of a right, left, and overall Petermann factor we can now identify and analyze physically and geometric distinct contributions from the right and left eigenmodes.

\section{Detailed evaluation}
\label{sec:detailed}
The general considerations of the previous section still leave open the question of how to determine the mode-nonorthogonality effects at and near EPs concretely and practically in specific systems. Given the mentioned singular nature of the Jordan decomposition, but physically consistent 
emerging response,  this calls for direct considerations  based on the Green's function itself. 

To do so, we start with a general scenario and decompose the given system into two coupled subsystems. This is written here as an $N\times N$ Hamiltonian
\begin{equation}\label{eq:Ham}
\Ham = \left(\begin{array}{cc}
\Ha & \cab   \\
\cba   & \Hb \\
\end{array}\right) ,
\end{equation}
where the $(N-\order) \times (N-\order)$ matrix $\Ha$ and the $\order \times \order$ matrix $\Hb$ describe two subsystems a and~b. The $(N-\order) \times \order$ matrix~$\cab$ and the $\order \times (N-\order)$ matrix $\cba$ describe a coupling of the two subsystems. 
It is convenient to write the vectors in the $N$-dimensional Hilbert space in a two-component representation
\begin{equation}%\label{eq:psi1}
	\ket{\Vector} = \left(\begin{array}{c}
		\ket{\vectora} \\
		\ket{\vectorb} \\
	\end{array}\right)	
\end{equation}
with $(N-\order)$-dimensional vector $\ket{\vectora}$ and $\order$-dimensional vector $\ket{\vectorb}$.
To avoid over-burdening the notation, we focus in this section on a specific EP with eigenvalue $E_l\equiv \lambda$ and also drop the index $l$ from the eigenvectors and the spectral strength. 
In the above two-component representation a right eigenvector of $\Ham$ is
\begin{equation}%\label{eq:psi1}
	\ket{\Rv} = \left(\begin{array}{c}
		\ket{\rva} \\
		\ket{\rvb} \\
	\end{array}\right)	
\end{equation}
and the corresponding eigenvalue equation is
\begin{eqnarray}\label{eq:tca}
\lambda\ket{\rva} & = & \Ha\ket{\rva} + \cab\ket{\rvb} \\
\label{eq:tcb}
\lambda\ket{\rvb} & = & \cba\ket{\rva} + \Hb\ket{\rvb} 
\end{eqnarray}
with, in general complex-valued, eigenvalue $\lambda$. We solve Eq.~(\ref{eq:tca}) for $\ket{\rva}$ using the Green's function 
\begin{equation}
\GFa(E) := (E\ident-\Ha)^{-1} 
\end{equation}
and obtain
\begin{equation}\label{eq:rafromrb}
\ket{\rva} = \GFa(\lambda)\cab\ket{\rvb} .
\end{equation}
Inserting this into Eq.~(\ref{eq:tcb}) gives a nonlinear eigenvalue equation
\begin{equation}\label{eq:neveq}
	\lambda\ket{\rvb} = \ham(\lambda)\ket{\rvb}  
\end{equation}
with the effective Hamiltonian 
\begin{equation}\label{eq:Heff}
	\ham(E) := \Hb + \cba\GFa(E)\cab .
\end{equation}
From the dimensionality-reduced eigenvalue equation~(\ref{eq:neveq}) all eigenvalues of the full Hamiltonian~(\ref{eq:Ham}) can be retrieved because of the nonlinearity with respect to the eigenvalues. The nonlinear eigenvalue equation~(\ref{eq:neveq}) is therefore also called isospectral reduction, see, e.g., Ref.~\cite{RPM20}. When an eigenvalue~$\lambda$ and eigenvector $\ket{\rvb}$ of the reduced system is determined, the corresponding eigenvector of the full Hamiltonian~$\Ham$ can be computed with the help of Eq.~(\ref{eq:rafromrb}) as
\begin{equation}\label{eq:Rr}
	\ket{\Rv} = \left(\begin{array}{c}
		\GFa(\lambda)\cab\ket{\rvb} \\
		\ket{\rvb} \\
	\end{array}\right) .
\end{equation}
The equivalent procedure based on a left eigenvector $\bra{\Lv}$ gives the same effective Hamiltonian~(\ref{eq:Heff}) with
\begin{equation}\label{eq:Ll}
	\bra{\Lv} = \left(\bra{\lvb}\cba\GFa(\lambda), \bra{\lvb}\right) .
\end{equation}
With a Schur decomposition, we can always bring a Hamiltonian by a unitary transformation into the form in Eq.~(\ref{eq:Ham}) with $W=0$ or $V=0$. In both cases, the effective Hamiltonian $\ham(E)$ equals $\Hb$. In the following, we consider the situation, where an EP$_\order$ of the $\order\times\order$ matrix $\Hb$ is also an EP$_\order$ of the full system.

Let us first define the $\order\times\order$ Green's function of the Hamiltonian~$\Hb$
\begin{equation}
	\GFb(E) := (E\ident-\Hb)^{-1} 
\end{equation}
and the $N\times N$ Green's function of the full Hamiltonian~(\ref{eq:Ham})
\begin{equation}
\GF(E) := (E\ident-\Ham)^{-1}  .
\end{equation}
To calculate the latter we have to solve 
\begin{equation}
\left(\begin{array}{cc}
\Gaa(E) & \Gab(E)  \\
\Gba(E) & \Gbb(E) \\
\end{array}\right) 
\left(\begin{array}{cc}
	E-\Ha & -\cab   \\
	-\cba   & E-\Hb \\
\end{array}\right) 
= \ident .
\end{equation}
A straightforward calculation shows that 
\begin{equation}
	\Gbb(E) = \GFb(E) .
\end{equation}
With this equation and the orthogonal projector onto subsystem $b$ 
\begin{equation}\label{eq:Pb}
\projectorb := \left(\begin{array}{cc}
0 & 0  \\
0 & \ident \\
\end{array}\right) 
\end{equation}
we obtain the relation between the two Green's functions
\begin{equation}\label{eq:PbGFPb}
\projectorb\GF(E)\projectorb = \GFb(E) .
\end{equation}
Note that the projector $\projectorb$ is directly related to the projectors $\hat P_l^{(R,L)}$ from the previous section used in Eqs.~(\ref{eq:kgeneralised3R}) and~(\ref{eq:kgeneralised3L}), namely $\projectorb = \hat P_l^{(R)}$ for $V=0$ and $\projectorb = \hat P_l^{(L)}$ for $W=0$. To describe both cases simultaneously, we write for the spectral response strengths defined in Eqs.~(\ref{eq:kgeneralised3R}) and (\ref{eq:kgeneralised3L}) in the following $\rcaintern$. 

Next we use the fact that the rank-1 matrix $\nilpotent^{\order-1}$ of the given EP can be written in terms of its right and left eigenvectors as in Eq.~(\ref{eq:nnl1})
\begin{equation}\label{eq:Wrca}
\nilpotent^{\order-1} = \rca\frac{\ket{\REP}\bra{\LEP}}{\normV{\REP} \normV{\LEP}} .
\end{equation}
Plugging this into the dominant contribution near the EP [Eq.~(\ref{eq:domicontri})] gives 
\begin{equation}\label{eq:GFRL}
	\GF(E) \sim \frac{\rca }{(\ev-\evEP)^\order} \frac{\ket{\REP}\bra{\LEP}}{\normV{\REP}\normV{\LEP}} .
\end{equation}
With Eq.~(\ref{eq:PbGFPb}) and $\ket{\rEPb} := \projectorb\ket{\REP}$ and $\ket{\lEPb} := \projectorb\ket{\LEP}$ we obtain
\begin{equation}\label{eq:gf1}
	\GFb(E) \sim \frac{\rca}{(\ev-\evEP)^\order} \frac{\ket{\rEPb}\bra{\lEPb}}{\normV{\REP}\normV{\LEP}}  .
\end{equation}
We write this in a form like in Eq.~(\ref{eq:GFRL})
\begin{equation}\label{eq:gf2}
	\GFb(E) \sim \frac{\rcaintern}{(\ev-\evEP)^\order} \frac{ \ket{\rEPb}\bra{\lEPb}}{\normV{\rEPb}\normV{\lEPb}} .
\end{equation}
The internal spectral response strength $\rcaintern$ in general differs from the spectral response strength of the full Hamiltonian, $\rca$.
Comparing Eqs.~(\ref{eq:gf1}) and (\ref{eq:gf2}) gives 
\begin{equation}
	\rca = \rcaintern \frac{\normV{\REP}\normV{\LEP}}{\normV{\projectorb\REP}\normV{\projectorb\LEP}} .
 \label{eq:Kintern}
\end{equation}
or with the EP Petermann factor in Eq.~(\ref{eq:kgeneralised})
\begin{equation}\label{eq:PFg}
	\PF^{(R,L)} = \left(\frac{\rca}{\rcaintern}\right)^2 = \left(\frac{\normV{\REP}\normV{\LEP}}{\normV{\projectorb\REP}\normV{\projectorb\LEP}}\right)^2 .
\end{equation}
For the Hamiltonian $\Ham$ in Eq.~(\ref{eq:Ham}) with the eigenvectors in Eqs.~(\ref{eq:Rr}) and~(\ref{eq:Ll}), it follows for $V=0$
\begin{equation}
	\PF^{(R)} = \left(1+\bra{\lEPb}\cba \GFa(\evEP)\GFa^\dagger(\evEP)\cba^\dagger\ket{\lEPb}\right)
 \label{eq:PFg2R}
\end{equation}
and for $W=0$
\begin{equation}
	\PF^{(L)} = \left(1+\bra{\rEPb}\cab^\dagger \GFa^\dagger(\evEP)\GFa(\evEP)\cab\ket{\rEPb}\right)
 \label{eq:PFg2L}
\end{equation}
where the vectors $\ket{\rEPb}$ and $\ket{\lEPb}$ are normalized to unity. Note that in general $\braket{\lEPb}{\rEPb} \neq 0$ even if $\braket{\LEP}{\REP} = 0$. This remarkable fact indicates that $\ket{\rEPb}$ and $\ket{\lEPb}$ are not the right and left EP-eigenvectors of $\Hb$. From Eqs.~(\ref{eq:PFg2R})-(\ref{eq:PFg2L}) or~(\ref{eq:PFg}) we can confirm that $\PF^{(R,L)} \geq 1$.

Beyond the advantages mentioned in the previous sections, $\PF^{(R,L)}$ and $\rcaintern$ offer the following additional benefits. The spectral response strength $\rca$ quantifies the spectral response to general perturbations of the full system expressed by an $N\times N$ perturbation matrix $\Hp$ in terms of the inequality~\cite{Wiersig22,Wiersig23b}
\begin{equation}\label{eq:specresponse}
	|\ev_j-\evEP|^\order \leq \varepsilon \normspec{\Hp}\,\rca  .
\end{equation}
The quantity $\rcaintern$ on the other hand describes the spectral response to perturbations of the subsystem b alone, expressed by the matrix $\hp = \projectorb\Hp\projectorb$,
\begin{equation}\label{eq:specresponse2}
	|\ev_j-\evEP|^\order \leq \varepsilon \normspec{\hp}\,\rcaintern  .
\end{equation}
In the same spirit, $\rca$ describes the steady-state intensity response of the full system to arbitrary harmonic excitation with strength $P > 0$ and frequency $\pf$~\cite{Wiersig22}
\begin{equation}\label{eq:rtoe}
	\normV{\psi} \leq P\, \frac{1}{|\hbar\pf-\evEP|^{\order}}\, \rca .
\end{equation}
$\rcaintern$ describes the intensity response of subsystem b if the excitation or detection is restricted to subsystem b,
\begin{equation}\label{eq:rtoe2}
	\normV{\projectorb\psi} \leq P\, \frac{1}{|\hbar\pf-\evEP|^{\order}}\, \rcaintern\ .
\end{equation}
Note that in the inequalities~(\ref{eq:specresponse2}) and~(\ref{eq:rtoe2}), one must choose either $\rca^{(R)}$ for $V=0$ and $\rca^{(L)}$ for $W=0$.

\subsection{Toy model}
To illustrate our results,  we consider the following $3\times 3$ toy-model Hamiltonian (see Ref.~\cite{Wiersig23b})
\begin{equation}\label{eq:H33}
\Ham = \left(\begin{array}{ccc}
\Ea & A & 0\\
0   & \Eb & B\\
0   & 0 & \Eb\\
\end{array}\right) ,
\end{equation}
with complex parameters $\Ea$, $\Eb$, $A \neq 0$, and $B \neq 0$. The eigenvalues of $\Ham$ are $\Ea$ and $\Eb$. The latter has an algebraic multiplicity of two. By comparing with Eq.~(\ref{eq:Ham}), we identify $\Ha = \Ea$, $W = (0,0)^\transpose$, $V = (A,0)$, and 
\begin{equation}
\Hb = \left(\begin{array}{cc}
\Eb & B\\
  0 & \Eb\\
\end{array}\right) .
\end{equation}
The response strength of $\Hb$ at the second-order EP with eigenvalue $\evEP = \Eb$ is $|B|$, see Ref.~\cite{Wiersig22}. We identify this here with the internal response strength, that is, $\rca^{(L)} = |B|$. With $\ket{\rEPb} = (1,0)^\transpose$, $\bra{\lEPb} = (0,1)$, and $\GFa(\ev) = (\ev-\Ea)^{-1}$ we obtain in a short calculation from Eq.~(\ref{eq:PFg2L}) for the EP Petermann factor
\begin{equation}
\PF^{(L)} = 1+\frac{|A|^2}{|\Eb-\Ea|^2}
\end{equation}
and consequently
\begin{equation}\label{eq:toyxi}
\rca = |B|\sqrt{1+\frac{|A|^2}{|\Eb-\Ea|^2}} \ . 
\end{equation}
This result for the spectral response strength agrees with the result in Ref.~\cite{Wiersig23b} based on a different, for this example more involved, approach.

The internal response strength $\rca^{(R)}$ is different. From Eq.~(\ref{eq:kgeneralised3R}) we obtain
\begin{equation}
\rca^{(R)} = \frac{|B|(|\Ea-\Eb|^2+|A|^2)}{\sqrt{|A|^2(|\Ea-\Eb|^2+|B|^2)+|\Ea-\Eb|^4}} ,
\label{eq:toyxir}
\end{equation}
and using Eqs.~(\ref{eq:kgeneralised}) and (\ref{eq:toyxi}) we conclude after a short calculation
\begin{align}
\PF^{(R)} &= 1+\frac{|AB|^2}{|\Ea-\Eb|^2(|\Ea-\Eb|^2+|A|^2)} .
\end{align}
As expected, $\PF^{(R)}$ and $\PF^{(L)}$  diverge if $\Ea = \Eb$, which results in a higher-order degeneracy, an EP$_3$ (the physical meaning of such a divergence is discussed in Ref.~\cite{Wiersig23b}). This is unless $A=0$, which results in a degeneracy of algebraic multiplicity~3 and geometric multiplicity 2 (instead of 1 for a generic EP), as covered in the recent article~\cite{BS25}.

The expression for $\PF^{(R)}$ is more complicated than $\PF^{(L)}$ because the eigenvalues in question are placed in the last block of the Schur decomposition. This is as expected, as the decomposition involves a Gram-Schmidt orthogonalization that starts with the other eigenvectors in the system. Simpler expressions would be obtained  by  reordering the procedure to give a diagonal $\Eb,\Eb,\Ea$, or by bringing the system into lower-triangular form. In particular, in our setting, the matrix
\begin{equation}
    \left(\begin{array}{ccc}
1 & A & 0\\
0   & 0 & B\\
0   & 0 & 0\\
\end{array}\right)
\end{equation}
(which applies without loss of generality with energy shifted and scaled so that $\Ea - \Eb = 1$)
can be unitarily transformed into a matrix of the form
\begin{equation}
    \left(\begin{array}{ccc}
0 & \frac{B(1+|A|^2)}{\sqrt{1+|A|^2+|AB|^2}} & \star\\
0   & 0 & \star\\
0   & 0 & 1\\
\end{array}\right),
\end{equation}
from which we directly recover Eq.~\eqref{eq:toyxir}; with $\star$ denoting terms that we do not need for its determination.

\section{Realistic example: Two waveguide-coupled microrings}
\label{sec:microrings}
\begin{figure}[t]
	\includegraphics[width=0.8\columnwidth]{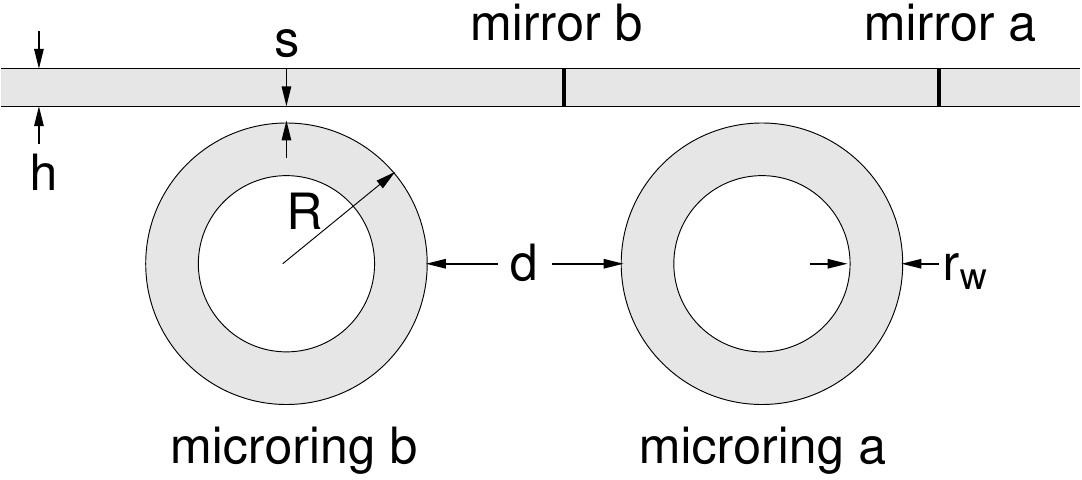}
	\caption{Illustration of the system consisting of two equal microrings coupled to an infinite waveguide, which contains two partial mirrors.}
	\label{fig:example}
\end{figure}

Having discussed a toy model in the previous section we now turn our attention to a more realistic example illustrated in Fig.~\ref{fig:example}, consisting of two microring cavities coupled to a waveguide, which contains two partial mirrors. Implementing second-order EPs in such microring-waveguide systems has been first proposed in Ref.~\cite{ZRK19} and shortly after experimentally realized in Ref.~\cite{SZM22}. The extension to higher-order EPs can be found in Ref.~\cite{KGK23}.
We use the traveling-wave basis with the counterclockwise propagating wave (CCW) in microring a, the clockwise propagating wave (CW) in microring a, the CCW propagating wave in microring b and the CW propagating wave in microring b. Following coupled-mode theory this system is described by the effective Hamiltonian
\begin{equation}\label{eq:Hexample}
\Ham = \left(\begin{array}{cccc}
\omega_{\text{a}} & r_{\text{a}}      & 0        & 0        \\
r_{\text{b}}		 & \omega_{\text{a}} & 0        & t        \\
t		 & 0        & \omega_{\text{b}} & r_{\text{b}}      \\
0		 & 0        & 0        & \omega_{\text{b}} \\
	\end{array}\right) 
\end{equation}
with the frequency $\omega_{\text{a}}$ ($\omega_{\text{b}}$ ) of microring a (b), reflectivity $r_{\text{a}}$ of mirror a and reflectivity $r_{\text{b}}$ and transmission coefficient $t$ of mirror b. 
This system is of the type of Eq.~(\ref{eq:Ham}), but with $W \neq 0 \neq V$. Hence,  Eqs.~(\ref{eq:PFg2R}) and~(\ref{eq:PFg2L}) do not apply to this more general system.

% t = 0
For the uncoupled case, $t = 0$, the subsystem~b is at an EP$_2$ with eigenvalue $\freqEP = \omega_{\text{b}}$ and spectral response strength~$|r_{\text{b}}|$. The latter we might identify naively with the internal response strength $\rcaintern = |r_{\text{b}}|$. However, we will see later in this section that this is not correct.

% eigenvalues and -vectors
For the coupled case, $t\neq 0$, the full system still has an EP$_2$, which can be verified by direct computation. The eigenvalues satisfy
\begin{equation}
	(\omega-\omega_{\text{b}})^2p(\omega) = 0
\end{equation}
with the second-order polynomial
\begin{equation}\label{eq:defp}
	p(\omega) := (\omega-\omega_{\text{a}})^2-r_{\text{a}} r_{\text{b}} .
\end{equation}
The eigenvalues are $\freqEP = \omega_{\text{b}}$ and $\omega_\pm = \omega_{\text{a}}\pm\sqrt{r_{\text{a}}r_{\text{b}}}$. In the following, we assume $\freqEP \neq \omega_\pm$ and $\freqEP \neq \omega_{\text{a}}$. The right and left EP eigenstates are
\begin{equation}\label{eq:exampleEPvec}
\ket{\REP} = \left(\begin{array}{c}
0 \\
0 \\
R \\
0 \\
\end{array}\right) 
\;\;\text{and}\;\;
\bra{\LEP} = \left(0,0,0,L\right) .
\end{equation}
While it is difficult to compute the Jordan vectors of an EP numerically in general scenarios, we can here compute them analytically. The result is
\begin{equation}\label{eq:exampleJordanR}
\ket{J_2} = \frac{R}{\alpha}\left(\begin{array}{c}
r_{\text{a}}t/p(\omega_{\text{b}}) \\
t(\omega_{\text{b}}-\omega_{\text{a}})/p(\omega_{\text{b}}) \\
\beta \\
1 \\
\end{array}\right) 
\end{equation}
and
\begin{equation}\label{eq:exampleJordanL}
\bra{\tilde{J}_1} = \frac{L}{\alpha} \left(\frac{t(\omega_{\text{b}}-\omega_{\text{a}})}{p(\omega_{\text{b}})},\frac{r_{\text{a}}t}{p(\omega_{\text{b}})},1,0\right) 
\end{equation}
with the abbreviation 
\begin{equation}\label{eq:alpha}
\alpha := r_{\text{b}}+\frac{t^2r_{\text{a}}}{p(\omega_{\text{b}})} \ .
\end{equation}
The biorthogonality relations~(\ref{eq:jordanbiorth}) require $LR = \alpha$ and
\begin{equation}
 \beta = -2r_{\text{a}}t^2\frac{\omega_{\text{b}}-\omega_{\text{a}}}{p^2(\omega_{\text{b}})} \ .   
\end{equation}
Plugging Eqs.~(\ref{eq:exampleEPvec})-(\ref{eq:alpha}) into Eq.~(\ref{eq:kgeneralised2}) gives the spectral response strength
\begin{equation}\label{eq:rcawaveguide}
\rca = |\alpha| = \left|r_{\text{b}}+\frac{t^2r_{\text{a}}}{p(\omega_{\text{b}})}\right| \ .
\end{equation}
This result can be confirmed with the method described in Ref.~\cite{Wiersig23b}. The subsystem a together with mirror b acts as a dispersive mirror, hence the reflectivity $r_{\text{b}}$ is replaced by $\alpha$. 

For the internal response we get according to Eqs.~(\ref{eq:kgeneralised3R}) and 
(\ref{eq:kgeneralised3L})
\begin{equation}
\rca^{(R,L)} = \frac{|t^2r_{\text{a}}+r_{\text{b}}p(\omega_{\text{b}})|}{\sqrt{|t|^2(|r_{\text{a}}|^2 + |\omega_{\text{b}}-\omega_{\text{a}}|^2) + |p(\omega_{\text{b}})|^2}} .
\end{equation}
It has to be emphasized that in this special case $\rca^{(R)} = \rca^{(L)}$ and that this result significantly differs from the naive guess $\rcaintern = |r_{\text{b}}|$ made at the beginning of this section.

The EP Petermann factor in Eq.~(\ref{eq:kgeneralised})  is then given by
\begin{equation}
	K^{(R,L)} = 1 + \frac{|t|^2(|r_{\text{a}}|^2 + |\omega_{\text{b}}-\omega_{\text{a}}|^2)}{|p(\omega_{\text{b}})|^2} .
\end{equation}
In the remaining part of this section we show that the effective Hamiltonian in Eq.~(\ref{eq:Hexample}) models a realistic photonic system. As the system is quasi-two-dimensional, Maxwell's equations reduce to the scalar mode equation
\begin{equation}
	\label{eq:ModeEq}
	\left(\Delta + k^2 n^2\right)\psi = 0
\end{equation}
with the complex wave number~$k=\omega/c$, the speed of light in vacuum~$c$, and the complex frequency~$\omega$. Using the radius~$R$ of the microrings as a characteristic length scale allows to rescale the frequency to a dimensionless complex frequency $\Omega=R\omega/c$. The wavefunction $\psi(x,y)$ describes either the electric or the magnetic field perpendicular to the cavity plane. In the following, we focus on the transverse magnetic polarization where the electric field is $\vec{E} = (0, 0,\text{Re}[\psi(x,y) e^{-i\omega t}])$.

The geometric parameters of the system, indicated in Fig.~\ref{fig:example}, are as follows: the thickness $r_{\text{w}} = 0.13R$ of the rings; the distance $s=0.133R$ between cavity and waveguide; the thickness $h=0.13R$ of the waveguide; the edge-to-edge distance $d=1.6R$ between the two cavities; the refractive index~$n=3.1$ of both microrings and waveguide. The two mirrors in the waveguide are implemented by locally changing the refractive index to $n=0.5+10i$ modeling gold. Mirror~b is placed symmetrically between the cavities and has width of~$d_{\text{b}} = 0.005R$. Mirror~a has a width of~$d_{\text{a}} = 0.06R$ and is placed such that cavity a is symmetrically between both mirrors.

\begin{figure}[tb]
	\begin{center}
		\includegraphics[]{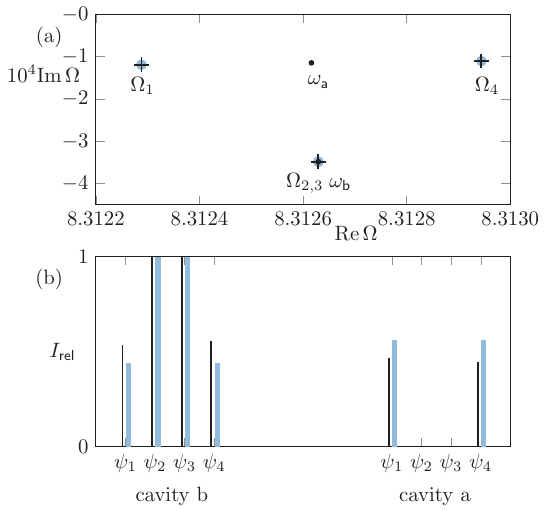}%
		\caption{(a) Complex frequencies $\Omega$ for the system shown in Fig.~\ref{fig:example}. Results from FEM simulations are shown in thick blue dots. Black pluses are the eigenvalues of the effective Hamiltonian~(\ref{eq:Hexample}). The parameters $\omega_\text{a}$ and $\omega_\text{b}$ are shown as thin black dots. (b) Relative intensity~$I_{\text{rel}}$ of the modes in cavity~a and cavity~b. Blue thick bars are FEM simulations (cf. Fig.~\ref{fig:SystemModes}) and thin black bars correspond to the eigenstates of the effective Hamiltonian~(\ref{eq:Hexample}).}
		\label{fig:SystemResults}
	\end{center}
\end{figure}

The software package JCMsuite based on the finite-element method (FEM)~\cite{PomBurZsc2007} is used to solve Eq.~\eqref{eq:ModeEq} for long-lived modes with $\Omega\approx8.31$. The numerical results are four modes $\psi_{1-4}$ with complex frequencies $\Omega_{1-4}$. However, as shown in Fig.~\ref{fig:SystemResults}(a) two of them have (almost) the same complex frequency $\Omega_2=\Omega_3$ and mode patterns $|\psi_{2}| = |\psi_{3}|$ indicating an EP of second order. We therefore refer to these two modes as $\Omega_{\text{EP}}$ and $\psi_{\text{EP}}$. 

\begin{figure}[tb]
	\begin{center}
		\includegraphics[]{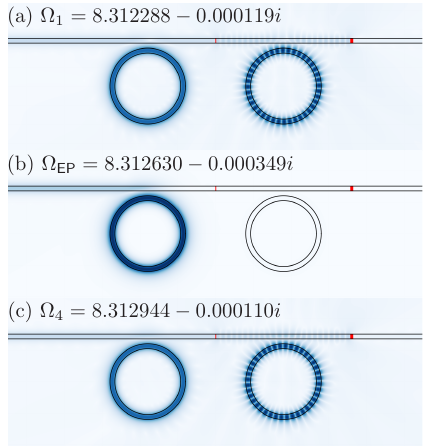}%
		\caption{Mode patterns from FEM simulations for the system of two waveguide-coupled microrings; cf. Fig.~\ref{fig:example}. Shown is $|\psi(x,y)|^{1/2}$ to enhance the visibility of small intensities. In the waveguide are two semi-transparent mirrors: Between the cavities is a thin mirror with width $d_{\text{b}} = 0.005R$ and right to the cavities is a thick mirror with width $d_{\text{a}} = 0.06R$. Both mirrors are indicated by red rectangles.}
		\label{fig:SystemModes}
	\end{center}
\end{figure}

As can be seen in Fig.~\ref{fig:SystemModes}, the EP mode $\psi_{\text{EP}}$ corresponds to a pure CCW wave in cavity b. The two non-EP modes $\psi_1$ and $\psi_4$ exhibit a field in both cavities; see also Fig.~\ref{fig:SystemResults}(b). Thus, the overlap integral
\begin{equation}
	S[\psi_i, \psi_j] = \frac{\int_M \psi_i^* \psi_j\,\text{d}^2r}{\sqrt{\int_M  |\psi_i|^2\,\text{d}^2r \int_M |\psi_j|^2\,\text{d}^2r}} ,
\end{equation}
with $M$ being the area of the two microrings, is finite for all combinations $i,j=1,...,4$. In particular, modes $\psi_1$ and $\psi_4$ both have an overlap with the EP mode $\psi_{\text{EP}}$ that is $|S|\approx 0.66$ while the overlap between mode $\psi_1$ and $\psi_4$ is $|S|\approx0.29$. In addition, the mode $\psi_{\text{EP}}$ shows the self-orthogonal feature with $|S[\psi_{\text{EP}}^*, \psi_{\text{EP}}]| \approx 6.9\times10^{-4}$.

Employing the fit procedure described in Appendix \ref{sec:AppFitProcedure} yields the parameters in the effective Hamiltonian \eqref{eq:Hexample} which correspond to the FEM-based results: In particular, $H$ has the eigenvalues $\Omega_{1-4}$ and the eigenvectors show the same relative intensity distribution among the microrings as $\psi_{1-4}$, see Fig.~\ref{fig:SystemResults}. In addition, the fitting procedure fixes the phase of the parameters in $H$ such that Eq.~\eqref{eq:rcawaveguide} can be used to calculate the response strength $\xi$ of the EP. A reference value $\xi_{\text{num}}$ for the response strength is obtained from the FEM simulations by the algorithm described in Ref.~\cite{KulWie2025}; see also Appendix~\ref{sec:AppXiCalc}. 

For different combinations $(d_{\text{b}}, d_{\text{a}})$ of mirror thicknesses the results are summarized in Tab.~\ref{tab:XiResults}. Varying the thickness of the mirrors does not change the fact that the system is at a second-order EP but it modifies the parameters in the Hamiltonian and therefore impacts the response strength via the mode non-orthogonality. In each case, the response strength calculated from Eq.~\eqref{eq:rcawaveguide} agrees much better with the numerical reference than the value $\xi_{\text{sub}} = |r_{\text{b}}|$ of the subsystem which neglects the presence of the mode $\psi_1$ and $\psi_4$. 
Based on the general insights obtained in the present work, the small remaining differences can be attributed to the mode-nonorthogonality with spectrally nearby, lower-$Q$ Fabry-P{\'e}rot modes formed by the two mirrors in the waveguide. 

\begin{table}[tb!]
	\begin{ruledtabular}
		\begin{tabular}{ccccc}
			$d_{\text{b}}/R$&  $d_{\text{a}}/R$& $10^3\xi$ & $10^3\xi_{\text{sub}}$ & $10^3\xi_{\text{num}}$ \\
			\hline
			$0.003$& $0.060$  & 1.683& 0.194& 1.738 \\
			$0.005$& $0.060$  & 1.055& 0.248& 1.182 \\
			$0.007$& $0.060$  & 0.795& 0.279& 0.968 \\
			$0.009$& $0.060$  & 0.657& 0.300& 0.861 \\
			$0.003$& $0.003$  & 2.183& 0.275& 1.738 \\
			$0.005$& $0.005$  & 1.359& 0.362& 1.182 \\
		\end{tabular}
	\end{ruledtabular}
	\caption{\label{tab:XiResults} Results for the spectral response strength for different combinations $(d_{\text{b}}, d_{\text{a}})$ of the thicknesses of the mirrors in the waveguide.}
\end{table}

\section{Conclusion}
\label{sec:conclusion}
In this work, we have generalized the concept of the Petermann factor, which was originally introduced for isolated resonances, to the extreme situation at a (potentially high-order) EP. This \emph{EP Petermann factor} is a measure of mode-nonorthogonality of generalized eigenspaces in the same way as the conventional Petermann factor measures the non-orthogonality of isolated resonances. 

We have linked this new quantity to the spectral response strength, which quantifies the spectral response of a non-Hermitian system with an EP in terms of frequency splitting caused by a perturbation and the intensity response to an external excitation, including the original context of excess quantum noise of bad-cavity lasers.

As a key step to identify the EP Petermann factor, we introduced the concept of \emph{internal spectral response strengths} that quantify the spectral response to subsystem perturbations and the intensity response to subsystem excitations. 
An important observation is that the EP subsystem can be constructed in two natural, generally inequivalent, ways, which relate to the generalized left and right eigenspaces associated with the EP. In the setting of dynamical response, the left eigenspace is known to relate to the source of the excitations, while the right eigenspace refers to the detection of the response. 
Practically, the corresponding EP Petermann factors then capture the difference in the spectral response of the total system relative to the response of the EP subsystem.
Therefore, these left and right EP Petermann factors could be determined experimentally by comparing the total response power \eqref{eq:pep} with the response powers \eqref{eq:pepr} and \eqref{eq:pepl} in the corresponding source and detection configurations. First experimental steps into determining the left and right eigenstates have been reported in Ref.~\cite{WA25}. Finally, these two Petermann factors  can be combined, via their geometric mean, into an overall Petermann factor, which itself again carries a well-defined geometric meaning. 

We furthermore showed how the right and left EP Petermann factors can serve to accurately calculate the overall response strength in model systems. 
We have illustrated this with a simple toy model and with a realistic photonic set consisting of two microrings coupled to a waveguide with embedded semitransparent mirrors. 

Our theory provides a comprehensive and unified characterization of non-Hermitian systems operating at EPs and sets the stage for the development of open systems with unique spectral responses, whether in terms of emission or reactions to static and dynamic perturbations.

Supplementary data tables and source code for the numerical experiments for this work can be found in the open-access data publication~\cite{KWSdata25}.

\acknowledgments 
Fruitful discussions with Subhajyoti Bid are acknowledged. We acknowledge support for the Book Processing Charge by the Open Access Publication Fund of Magdeburg University.

\begin{appendix}
\section{Derivation of the geometric expressions for the EP Petermann factor\label{app:kgeom}}
To see how the EP Petermann factor \eqref{eq:kgeneralised} can be expressed purely geometrically as given in Eqs.~\eqref{eq:explicitR} and ~\eqref{eq:explicitL},
we first determine the projections of the nilpotent part $\hat N_l$ onto the right and left generalized eigensubspaces.
Using the expressions \eqref{eq:nilpotent} for $\hat N_l$,  
\eqref{eq:explicitPR} for $\hat P_l^{(R)}$,
and 
\eqref{eq:explicitPL} for $\hat P_l^{(L)}$, the definitions of the inverse Gram matrices $A=B^{-1}$ and $\tilde A=\tilde B^{-1}$ with 
$    B_{km}=\langle J^{(l)}_k|J^{(l)}_m\rangle$,
    $ \tilde B_{km}=\langle \tilde J^{(l)}_k|\tilde J^{(l)}_m\rangle
    $, 
as well as
the generalized biorthogonality conditions \eqref{eq:jordanbiorth},
these projections are given as
\begin{align}
\hat N_l^{(R)}\equiv \hat P_l^{(R)} \hat N_l \hat P_l^{(R)}
&=
\sum_{km}A_{k+1,m}
|J^{(l)}_k\rangle\langle J^{(l)}_m|
,
\\
\hat N_l^{(L)}\equiv
\hat P_l^{(L)} \hat N_l \hat P_l^{(L)}
&=
\sum_{km}\tilde A_{k,m-1}
|\tilde J^{(l)}_k\rangle\langle \tilde J^{(l)}_m|,
\end{align}
where matrix elements with indices outside the index range $1,2,3,\ldots, n_l$ are taken to vanish.

Next, we evaluate powers of these matrices. Using the same properties as stated above, these are given by
\begin{align}
(\hat N_l^{(R)})^r
&=
\sum_{km}A_{k+r,m}
|J^{(l)}_k\rangle\langle J^{(l)}_m|
,
\\
(\hat N_l^{(L)})^r&=
\sum_{km}\tilde A_{k,m-r}
|\tilde J^{(l)}_k\rangle\langle \tilde J^{(l)}_m|.
\end{align}
We see that the matrices are nilpotent of degree $n_l$, while the power $r=n_l-1$ reduces to
\begin{align}
\hat X^{(R)}_l\equiv(\hat N_l^{(R)})^{n_l-1}
&=
|R_l\rangle\sum_{m}A_{n_l,m}
\langle J^{(l)}_m|
,
\\
\hat  X^{(L)}_l\equiv(\hat N_l^{(L)})^{n_l-1}&=
\sum_{k}\tilde A_{k,1}
|\tilde J^{(l)}_k\rangle\langle L_l|,
\end{align}
where we used the identifications 
$|J^{(l)}_1\rangle=|R_l\rangle$,
$\langle\tilde J^{(l)}_{n_l}|=\langle L_l|$.

Using a final time the mentioned properties, we furthermore find the identities
\begin{align}
\sum_{m}A_{n_l,m}\langle J^{(l)}_m|=\langle L_l|\hat P^{(R)}
,
\\
\sum_{k}\tilde A_{k,1}
|\tilde J^{(l)}_k\rangle =\hat P^{(L)}_l |R_l\rangle.
\end{align}
Therefore 
\begin{align}
\hat  X^{(R)}_l = |R_l\rangle\langle L_l| \hat P^{(R)}_l 
,
\\
\hat  X^{(L)}_l = \hat P^{(L)}_l |R_l\rangle\langle L_l| .
\end{align}
According to their definitions 
\eqref{eq:kgeneralised3R} and
\eqref{eq:kgeneralised3L},
the internal response strengths are given by
\begin{align}
    \left(\xi_l^{(R)}\right)^2
=\mbox{tr}\, \hat  X^{(R)\dagger}_l \hat  X^{(R)}_l
,
\\
    \left(\xi_l^{(L)}\right)^2
=\mbox{tr}\, \hat  X^{(L)\dagger}_l \hat  X^{(L)}_l.
\end{align}
Inserting here our expressions for  $\hat  X^{(R)}_l$ and  $\hat  X^{(L)}_l$, we obtain the internal spectral strengths in their compact form \eqref{eq:xiRcompact} and \eqref{eq:xiLcompact}.
The desired expressions ~\eqref{eq:explicitR} and ~\eqref{eq:explicitL} for the Petermann factor then follow by inserting these, along with Eq.~\eqref{eq:kgeneralised2}, into the definition \eqref{eq:kgeneralised}.

For completeness, we also state the projection rules for the idempotent part $\hat P_l$ of the generalized spectral decomposition     \eqref{eq:decnonh2}. These are given by
\begin{align}
\hat P_l \hat P_l^{(R)}&=\hat P_l^{(R)},\\
\hat P_l^{(R)}\hat P_l &=\hat P_l,\\
\hat P_l \hat P_l^{(L)}&=\hat P_l,\\
\hat P_l^{(L)}\hat P_l &=\hat P_l^{(L)},
\end{align}
and hence 
\begin{align}
\hat P_l^{(R)}\hat P_l \hat P_l^{(R)}&=\hat P_l^{(R)},\\
\hat P_l^{(L)}\hat P_l \hat P_l^{(L)}&=\hat P_l^{(L)}.
\end{align}
Furthermore, we observe the partial orthogonality conditions ($l\neq l'$)
\begin{align}
\hat P_{l'} \hat P_l^{(R)}&=0,\\
\hat P_{l}^{(L)} P_{l'} &=0,\\
\hat N_{l'} \hat P_l^{(R)}&=0,\\
\hat P_{l}^{(L)} N_{l'} &=0.
\end{align}
Starting again from the generalized spectral decomposition     \eqref{eq:decnonh2},  this also implies that the Hamiltonian becomes projected to
\begin{align}
\hat P_l^{(R)}\hat H \hat P_l^{(R)}=E_l \hat P_l^{(R)}+\hat N_l^{(R)},
\\
\hat P_l^{(L)}\hat H \hat P_l^{(L)}=E_l \hat P_l^{(L)}+\hat N_l^{(L)}.
\end{align}
This simple behavior underlines the privileged status of these two natural orthogonal projections.

\section{Proof of identity \eqref{eq:identity}}
\label{app:b}
To show identity \eqref{eq:identity} involving the root of unity $w=\exp(2\pi i /n_l)$ and an arbitrary complex constant $a$,
we expand each term into a geometric series, interchange summations, and utilize 
\begin{equation}
    \sum_{l'=1}^{n_l} w^{kl'}=
     \begin{cases}
n_l & \text{if } k \mod n_l = 0 \\
0 & \text{otherwise}
\end{cases} ,
% n_l \delta_{k\mod n_l,0}
\end{equation}
after which we can resum the series.
For $|a|<1$, these steps take the form
\begin{align}
\sum_{l'=1}^{n_l}\frac{1}{aw^{l'}-1}&=
-\sum_{k=0}^\infty\sum_{l'=1}^{n_l}a^kw^{kl'}
\nonumber\\
&=
-\sum_{k'=0}^\infty a^{n_lk'}n_l
\nonumber\\
&= 
\frac{n_l}{a^{n_l}-1}.
\end{align}
For $|a|>1$, we have
\begin{align}
\sum_{l'=1}^{n_l}\frac{1}{aw^{l'}-1}&=
\sum_{l'=1}^{n_l}\frac{1}{aw^{l'}}\frac{1}{1-a^{-1}w^{-l'}}
\nonumber\\
&=
\sum_{k=1}^\infty\sum_{l'=1}^{n_l}a^{-k}w^{-kl'}
\nonumber\\
&=
\sum_{k'=1}^\infty a^{-n_lk'}n_l
\nonumber\\
&= a^{-n_l}
\frac{n_l}{1-a^{-n_l}},
\end{align}
from which the same result follows.
For the edge case $|a|=1$, the identity correctly recovers the poles at $a=w^{-l'}$, as $Q(a)\equiv a^{n_l}-1=\prod_{l'=1}^{n_l}(a-w^{-l'})$.

Furthermore, reading the identity backwards, it can be verified for all $a$ by using the
partial fraction decomposition
\begin{equation}
\frac{P(a)}{Q(a)}=\sum_{l'=1}^{n_l}\frac{P(w^{-l'})}{Q'(w^{-l'})}\frac{1}{a-w^{-l'}}
\end{equation} 
of the stated rational function in terms of the simple roots $w^{-l'}$ of $Q(a)$,
where we set  $P(a)=n_l$ and use $Q'(w^{-l'})=n_lw^{-n_l l'}=n_lw^{l'}$.

\section{Fitting procedure \label{sec:AppFitProcedure}}
The goal in this Appendix is to determine all parameters in the effective Hamiltonian \eqref{eq:Hexample} from the numerically computable quantities $\Omega$ and $\psi$ that solve the mode equation \eqref{eq:ModeEq}. This is a non-trivial task as all parameters in $H$ are complex such that their phase cannot be fixed by comparing the intensity pattern of the eigenstates in the cavities. However, the phase of the parameters enter into the calculation for the response strength and therefore needs to be determined correctly. To do so, we employ a two-step fitting procedure. First, proper basis wavefunctions are constructed to project the modes $\psi_{1-4}$ into a $\mathbb{C}^4$ vector space. Second, a direct comparison of the projected modes to the eigenstates of the effective Hamiltonian~\eqref{eq:Hexample} allows for a unique fit of the parameters.

\subsection{Auxiliary systems for basis modes}
The effective Hamiltonian \eqref{eq:Hexample} is written in a particular basis of a $\mathbb{C}^4$ vector space where one identifies the first two entries with CCW and CW propagating waves in cavity a and the last two entries with CCW and CW propagating waves in cavity b. The goal of this section is to construct similar basis states $\phi_{1-4}$ from FEM simulations that can be used to express the system eigenstates $\psi_{1-4}$ accordingly in a $\mathbb{C}^4$ with the same interpretation as for the effective Hamiltonian.

\begin{figure}[tb]
	\begin{center}
		\includegraphics[]{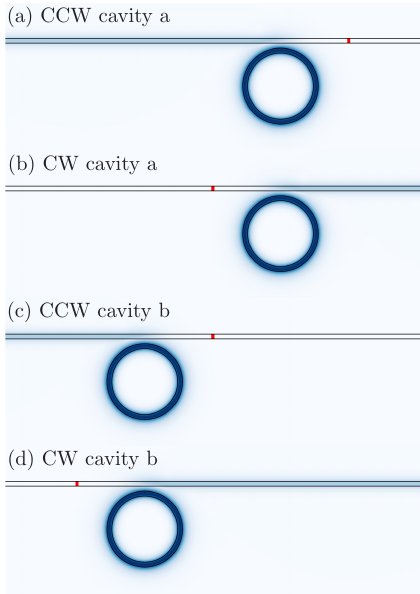}%
		\caption{Mode patterns $|\phi_{1-4}(x,y)|^{1/2}$ for different configurations of a single cavity coupled to a waveguide with a mirror of width $d=0.06R$. The mirror is indicated by a red rectangle.}
		\label{fig:BasisModes}
	\end{center}
\end{figure}

To do so, we design artificial systems that support pure CW and CCW propagating waves in each cavity. This is achieved by placing a single cavity--either cavity a or b--next to a waveguide. A single mirror of width $d=0.06R$ in the waveguide is then used to tune the system to an exceptional point via mirror-induced asymmetric backscattering~\cite{Wiersig18b,ZRK19}. Thus, the mode is either a pure CCW or CW propagating wave depending on the mirror being left or right to the cavity. This mode is used as the respective basis state $\phi_i$ as shown in Fig.~\ref{fig:BasisModes}. 

With the fixed set of basis states $\phi_{1-4}$ each mode of the system $\psi_j$ can be mapped to a vector $a^{(j)}\in\mathbb{C}^4$ with components calculated from the overlap integral 
\begin{align}
	a_i^{(j)} = S[\phi_i, \psi_j]
\end{align}
where the integration area $M$ is the area of the cavity from which the basis mode is constructed. Each mode $\psi_j$, in principle, can be multiplied by an arbitrary phase. To remove this ambiguity, the phase of each vector $a^{(j)}$ is specified to fulfill 
\begin{equation}
	\text{Arg}\left(a_3^{(j)}\right) = 0.
\end{equation} 
Thus, a unique representation of the modes $\psi_{1-4}$ as a four-dimensional vector is defined. Each component of the vector yields the same interpretation as the basis states of the effective Hamiltonian~\eqref{eq:Hexample}.

\subsection{Calculation of the effective Hamiltonian}
Next we determine the parameters in the effective Hamiltonian~\eqref{eq:Hexample} from the calculated frequencies $\Omega_i$ and the mode's representation as vector $a^{(j)}\in\mathbb{C}^4$.
From the structure of the effective Hamiltonian it follows that the diagonal elements can be calculated as
\begin{equation}
	\omega_{\text{a}} = \frac{\Omega_1 + \Omega_4}{2} \quad\text{and}\quad \omega_{\text{b}} = \Omega_\text{EP}.
\end{equation}
In addition, a comparison of non-degenerate eigenvalues of $H$ with the complex frequencies yields the relation
\begin{equation}
	\label{eq:sqrtRaRb}
	\sqrt{r_{\text{a}} r_{\text{b}}} = \frac{\Omega_4 - \Omega_1}{2}.
\end{equation}
Thus, only one of the parameters $r_{\text{a}}$ and $r_{\text{b}}$ needs to be determined and the other is fixed by Eq.~(\ref{eq:sqrtRaRb}). In the following, we choose $t$ and $r_{\text{b}}$ to be specified for the Hamiltonian $H$. Consequently, $H$ will have the eigenvalues $\Omega_1, \Omega_2$, and $\Omega_{\text{EP}}$ by construction. To determine the parameters $t$ and $r_{\text{b}}$ a cost function
\begin{align}
	f(t, r_{\text{b}}) = \sum_{i,j}\left\vert v_i^{(j)} - a_i^{(j)}\right\vert^2
\end{align}
is used that compares the eigenstates $v^{(j)}$ of the Hamiltonian to the previously calculated representation of the mode $a^{(j)}$ as complex vectors. The minimization of the cost function is done with a standard algorithm. The obtained values $t=10^{-4}(4.674614-2.915369i)$ and $r_{\text{b}}=10^{-4}(-1.541910-1.943754i)$ are unique in the sense that different initial guesses for the cost function optimization lead to the same parameter values.

\section{Response strength computation \label{sec:AppXiCalc}}
A reference value of the response strength is obtained directly from FEM simulations by the algorithm described in Ref.~\cite{KulWie2025}. We here summarize the algorithm for the system shown in Fig.~\ref{fig:SystemModes}. First, the Petermann factor is calculated numerically as
\begin{equation}
	K_l = \frac{\left\vert\int_M |\psi_l|^2 \text{d}r \right\vert^2}{\left\vert\int_M \psi_l^2 \text{d}r  \right\vert^2}
\end{equation}
where $M$ is the area of both microrings. Since the system is numerically close but not exactly at the EP the value for both modes $K_2$ and $K_3$ is finite. In combination with the numerically small difference between $\Omega_2$ and  $\Omega_3$ the numerical response strength of the EP of order $N=2$ is given by
\begin{equation}
	\xi_{\text{num}, l} = N\sqrt{K_l} |\Omega_l - \Omega_{\text{EP}}|^{N-1} 
\end{equation}
where $\Omega_{\text{EP}}=(\Omega_2+\Omega_3)/2$. For the two modes of interest $\xi_{\text{num}, 2}$ and $\xi_{\text{num}, 3}$ differ less than $10^{-6}$ yielding $\xi_{\text{num}} =(\xi_{\text{num}, 2} + \xi_{\text{num}, 3})/2 = 0.001182$.

\end{appendix}

%\bibliography{fg4,extern}

\end{document}